\documentclass[11pt,a4paper]{scrartcl}

\usepackage{CLICdp}
\pdfoutput=1
\usepackage{CLICdp_definitions}

\newcommand{\Delphes}{\textsc{Delphes}\xspace}
\newcommand{\DDHEP}{\textsc{DD4Hep}\xspace}
\usepackage{multirow}
\usepackage{listings}

\title{A \Delphes card for the CLIC detector}

\clicdpnote{2018}{007}  

\date{26 September 2019}

\addauthor{E.~Leogrande}{\institute{1}}
\addauthor{P.~Roloff}{\institute{1}}
\addauthor{U.~Schnoor\thanks{ulrike.schnoor@cern.ch}}{\institute{1}}
\addauthor{M.~Weber}{\institute{1}}
\addinstitute{1}{CERN, Switzerland}

\abstract{The Compact Linear Collider, CLIC, is a multi-TeV
  electron-positron collider proposed for construction at CERN.
  A detector model, CLICdet, that is suited for the experimental conditions at
  CLIC and is based on realistic  performance, has been developed.
  This paper describes the implementation of CLICdet in a fast
  simulation tool for particle physics collider experiments, \Delphes.
  The geometry of the detector concept as well as
  performance
  parameters extracted from full
  simulation studies are implemented in \Delphes parameter cards for
  CLICdet.
  Jet reconstruction for electron-positron colliders is added to the \Delphes
  analysis chain.
  Parameters for using  \Delphes to simulate the detector effects of
  CLICdet
  are provided in three parameter cards, one for each
  energy stage of CLIC.
  The effects of beam-induced background at the higher-energy stages of
  CLIC are also incorporated.
  The results from the fast simulation with \Delphes are validated
  with respect to full detector simulation in a number of
  relevant processes.
}

\titlecomment{This work was carried out in the framework of the CLICdp Collaboration} 

\graphicspath{ {./logos/}{./figures/} }

\begin{document}

\titlepage

\section{DELPHES for fast simulation of detector effects at CLIC}
The detector model for CLIC, CLICdet, has been defined and
described in~\cite{AlipourTehrani:2254048} and its performance is discussed in detail in~\cite{DetectorPerformance2018}.
Full simulation collision events can be obtained by running
the simulation and reconstruction chain on hadron-level Monte Carlo events generated e.g. with \textsc{Whizard}~\cite{Kilian:2007gr} and \textsc{Pythia6}~\cite{Sjostrand:2006za}.
Such a full detector simulation is performed for signal and background events for most of the studies carried out in the
framework of the CLICdp collaboration,
e.g.~\cite{Abramowicz:2210491,Abramowicz:2018rjq}.
 In order to optimise the design of the detector and to understand fully its impact on the proposed physics programme, the full detector simulation and event reconstruction framework is essential. However, for many studies including some of those of an exploratory nature or where the relative sensitivity to physics parameters is paramount, a faster simulation tool is extremely useful.

The fast simulation framework \Delphes~\cite{deFavereau2014,Selvaggi:2014mya,Mertens:2015kba} is a
modular system  which emulates the impact that the full detector simulation will  have on hadron-level simulation of collision events.  These include the  effects of detector geometry, detector
response and resolution as well as reconstruction performance of
physics objects.
It can be used to approximate those kinematic distributions for which the detector effects are well described by a parametrisation.
The simulation with \Delphes requires much less CPU time
than a full detector simulation which models the interactions of particles with the material of the detector.
According to the specific detector, an input parameter card must
be provided.
This document describes the parameter card developed for the CLICdet
detector model.
It has been used for several studies carried out in the framework of the recent report on the CLIC potential for new physics~\cite{BSMYR}.

A generic layout of a general-purpose detector for particle collider experiments is
implemented in \Delphes, of which some geometry parameters can be modified according
to the given detector model.
In addition, the performance of identification and reconstruction based on
certain objects and particle types can be defined in the detector
card.
The geometry and performance parameters inserted into the CLICdet card are described in Sec.~\ref{sec:parameters}.
The order in which detector effects are applied to generator level objects is described in Sec.~\ref{sec:applicationchain}.
Until recently, \Delphes has been used predominantly for studies at hadron colliders, in
particular the LHC.
Jet clustering at \epem colliders offers additional features which have been implemented into the \Delphes framework in the context
of the development of the parameter card.
This is described in Sec.~\ref{sec:jetclustering}.
Sec.~\ref{sec:validation} describes the validation of the \Delphes card, comparing its results for important physics processes with full detector simulation. Additional validation plots can be found in Appendix~\ref{app:extraplots}.
The parameter card has been merged into the central \Delphes repository
and can be accessed at \cite{Delphesgithub}.
User instructions are given in Appendix~\ref{sec:instructions}.

\section{Performance parameters of the CLICdet detector}\label{sec:parameters}
\subsection{The CLICdet detector}

The current detector model for CLIC, CLICdet, is adapted to the high-energy \epem collision environment at
CLIC and optimised for particle flow reconstruction.
The CLICdet concept was developed based on the lessons learned from
the previous detector models for CLIC: CLIC\_ILD and CLIC\_SiD~\cite{cdrvol2}.

The CLICdet model comprises a low-mass silicon vertex and
tracking system with a central barrel and forward disks.
CLICdet is optimised for particle flow in particular by using highly
granular calorimetry based on
silicon-tungsten electromagnetic calorimeters and scintillator-steel
hadronic calorimeters.
A superconducting solenoid surrounding the calorimeters provides a
magnetic field of 4\,T.
Beyond the solenoid, an iron yoke is interleaved with muon chambers.
In the forward region, close to the beam pipe, CLICdet contains
two specialised forward electromagnetic calorimeters to measure
luminosity and beam properties as well as to extend the acceptance.

The two previous models differ from CLICdet in the following main
aspects\cite{AlipourTehrani:2254048}:
The CLIC\_SiD concept uses a 5\,T magnetic field and a smaller radius
tracking system.
CLIC\_ILD uses a TPC as tracking system.
In both cases, tungsten is used as absorber in the barrel of the hadronic calorimeter.
While the CLIC \Delphes card is based on the CLICdet performance, some
of the validation studies for which no CLICdet simulation was available
are based on one of the previous detector concepts.

The right-handed coordinate system has its origin at the centre of the detector,
the $z$ axis along the detector axis close to the beam direction, and the $y$ axis pointing
upwards.
While most studies of $\epem$ collisions employ the polar angle
$\theta$ coordinate, defined with respect to the positive $z$
axis, this note mainly uses the pseudorapidity as it is the standard for 
\Delphes.
The pseudorapidity $\eta$ is defined as $\eta = -
\ln{[\tan{(\theta/2)}]}$ and the distance $\Delta R$ in the
pseudorapidity-azimuth plane is defined as $\Delta R = \sqrt{\Delta
  \eta^2 + \Delta \phi^2}$.

Under the baseline staging scenario, CLIC is expected to deliver 1\,ab$^{-1}$ of luminosity at $\sqrt{s}$\,=\,380\,GeV,
2.5\,ab$^{-1}$ at $\sqrt{s}$\,=\,1.5\,TeV, and 5\,ab$^{-1}$ at
$\sqrt{s}$\,=\,3\,TeV~\cite{Roloff:2645352}.
The luminosity will be shared evenly between an electron beam
polarisation of $P(\Pem)=-80\,\%$ and $+80\,\%$ at the first stage. At
the higher energy stages, 80\,\% (20\,\%) of the luminosity will be collected
with $P(\Pem)=-80\,\%$ ($P(\Pem)=+80\,\%$).

\subsection{Performance parameters for the \Delphes card}
The performance of CLICdet has  been thoroughly evaluated using full detector simulation in~\cite{DetectorPerformance2018}.
The detector geometry of CLICdet for the full simulation is implemented within
the \DDHEP framework~\cite{dd4hep} and simulated in
\textsc{Geant4}~\cite{AGOSTINELLI2003250,1610988,ALLISON2016186}.
The reconstruction software is implemented in the Linear Collider
framework MARLIN~\cite{MarlinLCCD}.
Track reconstruction is performed with the Conformal Tracking method~\cite{Leogrande:2630512}.
Based on the reconstructed tracks and calorimeter deposits, particle
flow clustering is performed using PandoraPFA
\cite{THOMSON200925,Marshall:2012ry,Marshall:2015rfa}.
Flavour tagging is applied using the \textsc{LCFIPlus} package~\cite{Suehara:2015ura}.

Three cards for the three energy stages of CLIC are provided in \Delphes.
Performance parameters for the CLICdet \Delphes cards are taken from studies for CLICdet described and referenced in the following.

\subsubsection{Track reconstruction performance}\label{sec:trackingperf}
The tracking efficiency is based on studies of the conformal tracking algorithm in~\cite[Fig. 33 and 35]{DetectorPerformance2018}.
Charged hadron efficiencies are based on the results for pions.
Efficiencies for muon tracks, electron tracks, and charged hadron
tracks are provided as a function of energy and pseudorapidity.
In addition, the momentum resolution from tracking is applied
according to a log-normal distribution with the resolution
given as $\frac{\Delta \pT}{\pT} = a \oplus
b/(p \sin{\theta})$ binned in pseudorapidity.
The parameters used in the CLICdet cards are derived from~\cite[Sec. 4.2.1, Fig. 31, 32]{DetectorPerformance2018}.

\subsubsection{Calorimeter performance}\label{sec:caloreso}
\Delphes takes into account the granularity of the calorimeter in terms of pseudorapidity and azimuthal angle.
Based on the geometry of the calorimeters  described in~\cite{AlipourTehrani:2254048}, the $\Delta \eta$ and $\Delta \phi$ segmentations implemented in the CLICdet cards are listed in Table~\ref{tab:calo}.

\begin{table}[htbp]
  \centering
  \caption{Calorimeter segmentations based on the CLICdet calorimeter cell sizes converted to $\Delta \phi  \times \Delta \eta$.}
  \label{tab:calo}
  \begin{tabular}[ht]{lcccc}\toprule
Part         & $\eta_{max}$ & cell size [mm$\times$mm] & $\Delta \phi [^{\circ}]$ & $\Delta \eta$ \\
\midrule
ECAL barrel  &    1.2 &              5 $\times$ 5 &        0.2 & 0.003 \\
ECAL endcaps &    2.5 &              5 $\times$ 5 &         0.8 & 0.02 \\
ECAL plug    &    3.0 &              5 $\times$ 5 &         1.0 & 0.02  \\
\midrule
HCAL barrel  &    0.8 &             30 $\times$ 30 &        1 & 0.02 \\
HCAL ring    &    0.9 &             30 $\times$ 30 &        1 & 0.02 \\
HCAL endcaps &    3.5 &             30 $\times$ 30 &        6 &  0.1 \\
\bottomrule
  \end{tabular}
\end{table}

The energy resolution in the calorimeters is implemented in the
\Delphes cards in terms of  absolute value,  $  \Delta E = \sqrt{ n^{2} + s^{2} E + c^{2} E^{2}} $
  with noise term $n$, stochastic term $s$, and constant term $c$.
In the CLICdet cards, separate resolution parameters are provided for the inner barrel, barrel, transition region and endcaps based on the results from~\cite[Fig. 37]{DetectorPerformance2018}.
The resolution from neutral kaons is used for the HCAL, while the ECAL
resolution is determined from photons.

\subsubsection{Isolated leptons and photons}\label{sec:isolatedobjects}
  Electron, photon, and muon candidates are selected from Particle
    Flow objects.
  Identification efficiencies are then applied to mimic the behavior
  found in full
  simulation~\cite{Weber:2648827},
  which includes the effects of track reconstruction and particle flow analysis:
  for electrons with $E>3$\,GeV, muons with $E>2$\,GeV, and photons
  with $E>2$\,GeV, the identification efficiencies are given as a function of energy and pseudo\-rapidity~\cite[Fig. 38-40]{DetectorPerformance2018}.

The isolation of an electron, muon or photon is determined according to the jet content
in a cone of $\Delta R =0.1$ surrounding the object.
It is considered as isolated if the ratio of the transverse
momentum sum of the cone to the transverse momentum of the isolated object does not exceed 0.2.
These settings are adapted from isolation criteria used in various full-simulation studies with CLIC detector concepts, e.g.~\cite{Redford:1690648}.

\subsubsection{Heavy flavour tagging}\label{sec:btagging}
Three working points for tagging of $b$ jets with efficiencies of
50\,\%, 70\,\%, and 90\,\% are provided in the CLICdet cards.
 The corresponding mis-tagging efficiencies for light flavour and
 charm jets are
 extracted from the double spiral option of the vertex endcaps in~\cite{AlipourTehrani:1742993} in bins of pseudorapidity and energy.

The $c$-tagging performance of CLICdet described
in~\cite[Sec.4.2.6]{DetectorPerformance2018} has not been implemented yet in the CLICdet \Delphes cards.
 
 \subsubsection{Tau lepton tagging}\label{sec:tautagging}
 The $\tau$ tagging efficiencies for the CLIC TauFinder algorithm have
 been studied in~\cite{Muennich:1443551}.
 The CLICdet \Delphes cards
 use efficiencies averaged for the three processes studied there,
 resulting in the efficiencies reported in
 Table~\ref{tab:tautaggingeff}.
 Jets are tagged as $\tau$ jets according to these efficiencies, using
 a mis-identification rate of 3\,\%.
 As the TauFinder uses a relatively simple, cut-based approach it is
 expected that its performance can be significantly improved with
 respect to these numbers by introducing a multivariate analysis.
 \begin{table}[htbp]
  \centering
  \caption{Efficiencies of the TauFinder for linear collider detectors
    adapted from~\cite{Muennich:1443551}.}
  \label{tab:tautaggingeff}
    \begin{tabular}[ht]{lccccccc}\toprule
$p_{T}(\tau)$ [GeV] & $\ge 5 $ & $\ge 12.5 $ & $\ge 25$ & $\ge 50$ & $\ge 75$ & $\ge 125$ & $\ge 250$      \\      \midrule

$\epsilon$ & 0.84 & 0.79 & 0.74& 0.66 & 0.61& 0.51& 0.36      \\\bottomrule
      
    \end{tabular}
\end{table}
\subsection{Effect of beam-induced backgrounds}\label{sec:gammagammahadrons}
The high bunch charge density at CLIC leads to strong radiation from the
incoming beams due to their interaction with the field of the other
beam.
This results in beam-induced background processes, e.g. \gghadrons.
To account for its effects, a jet energy smearing is applied.
It should be noted that the current implementation of the jet energy
smearing assumes massless jets.

In the reconstruction, timing information of the energy deposits in
the detector is used to reduce the impact
of the \gghadrons background. At the 380\,GeV energy
stage, this allows almost complete mitigation of the effect~\cite{Brondolin:2645355}.
At the higher energy stages, some impact remains despite the timing cuts.
It is mimicked by applying an additional smearing of the jet energy.
The impact of \gghadrons background on the jet energy resolution depends on the
centre-of-mass energy, being  most
severe at $\sqrt{s}\,=$\,3\,TeV at CLIC. Therefore,  separate cards are supplied for
the three stages of CLIC, as given in Appendix~\ref{sec:instructions}.
For the energy stage at $\sqrt{s} \approx$\,380\,GeV, denoted as Stage
1, no jet energy
smearing is applied.
The jet energy is smeared according to the corresponding resolutions at Stages 2
($\sqrt{s} \approx$\,1.5\,TeV) and 3 ($\sqrt{s} \approx$\,3\,TeV).

\section{Fast simulation and reconstruction}
\subsection{Application chain}\label{sec:applicationchain}
The various components (\Delphes modules) are called in the CLICdet card in the following order:
Particles are propagated through the magnetic field of 4\,T with the \texttt{ParticlePropagator} module.
In the next step, charged hadrons, electrons, and muons are retained or rejected based on random numbers drawn according to the tracking efficiencies.
Their momenta are then smeared according to the corresponding momentum resolutions.
The tracking efficiencies as well as the momentum resolutions are
given as a function of pseudorapidity and transverse momentum, see Sec.~\ref{sec:trackingperf}.
Energy deposits in the electromagnetic and hadronic calorimeters are then generated from the given expected energy fractions, taking into account the granularity of the calorimeter towers and the resolutions described in Sec.~\ref{sec:caloreso}.
A particle flow algorithm is applied to the tracks and calorimeter deposits using the \texttt{EFlowMerger} module.
Next, photon, charged hadron, electron, and muon reconstruction efficiencies and isolation efficiencies are applied according to Sec.~\ref{sec:isolatedobjects}.
To prepare for the jet clustering, isolated electrons, muons, and
photons are removed from the list of objects passed to the jet finder
using an \texttt{EFlowFilter} module.
Jet clustering with the VLC algorithm is applied to the remaining objects as described in Sec.~\ref{sec:jetclustering}.
In the CLICdet \Delphes cards for the higher energy stages (Stages 2 and 3, $\sqrt{s}>1$\,TeV), the effect of \gghadrons background is mimicked by applying a \texttt{JetMomentumSmearing} module (see Sec.~\ref{sec:gammagammahadrons}).
Finally, the \texttt{JetFlavorAssociation}, \texttt{BTagging} and \texttt{TauTagging} modules are run to first find the true flavour of the jet and then to apply tagging efficiencies as well as mis-tagging efficiencies for $b$-jets and hadronic $\tau$ candidates. Three working points for $b$ tagging with tagging efficiencies of 50\,\%, 70\,\%, and 90\,\% are provided in bins of pseudorapidity and energy (Sec.~\ref{sec:btagging}). Tagging efficiencies for hadronic $\tau$s are provided in bins of transverse momentum (Sec.~\ref{sec:tautagging}).

\subsection{Linear collider jet clustering}\label{sec:jetclustering}
Due to the lower levels of background at lepton colliders,
rather large cone size parameters $R$ can be used for jet clustering compared to typical values
at hadron colliders, leading to better
mass resolutions for hadronic resonances. 

 However, using larger $R$ implies that the jets can have a substructure of smaller $R$ jets. At analysis level, this is usually crucial information, e.g. for suppression of backgrounds with a  different jet multiplicity or in order to use the correct objects for invariant masses. 
 Therefore, jet clustering is often run in "exclusive mode"~\cite{Catani:1993hr} as implemented in Fastjet~\cite[Sec. 3.3.2]{Cacciari:2011ma}, with a fixed number of jets. This forces the algorithm to stop when a certain number of jets has been clustered, thus revealing the substructure information directly.

A suitable jet clustering algorithm for high-energy lepton collisions is the Valencia Linear Collider (VLC) algorithm~\cite{Boronat:2014hva,Boronat:2016tgd}. It is a sequential recombination algorithm adapted to the linear collider environment.

The VLC algorithm as implemented in
FastJet-contribs~\cite{fastjetcontribs}, the functionalities  for
exclusive jet clustering, and the related observables have been added
to the \Delphes framework as part of this work.
In the CLICdet cards, the VLC is applied with $\beta = \gamma = 1.0$ and for all combinations of $R = 0.5, 0.7, 1.0, 1.2,1.5$ and exclusive number of jets $N=2,3,4,5,6$ in order to accommodate a range of jet multiplicities that might be useful for the analyses conducted with these cards.
A large value of $R$ should be chosen  for the first centre-of-mass
stage of CLIC and slightly smaller $R$ for higher centre-of-mass due
to higher levels of beam-induced background.

\section{Validation of the \Delphes simulation}\label{sec:validation}
The performance of the \Delphes cards is validated for the three energy stages of CLIC separately by comparing various kinematic observables in important physics processes obtained with the corresponding CLICdet \Delphes card with full simulation based on the CLIC\_ILD, CLIC\_SiD, and CLICdet models.

Studies in the CDR~\cite{cdrvol2} use the CLIC\_ILD and CLIC\_SiD models.
Therefore, large event samples are available for these models and hence part of the validation uses CLIC\_ILD or CLIC\_SiD detector simulation.
The performance of these detector models is similar to that of
CLICdet for the relevant aspects.

The validation is based on the  processes summarised in
Table~\ref{tab:validationsamplse}.
For each of the samples, the CLICdet \Delphes card corresponding to the CLIC
energy stage has been used, i.e.  for Stage \texttt{N} the card named \texttt{delphes\_card\_CLICdet\_StageN.tcl}.

\begin{table}[htbp]
  \centering
  \caption{Overview of the samples used for validation of the CLICdet
    \Delphes cards.}
  \label{tab:validationsamplse}
  \begin{tabular}[ht]{lccclll}\toprule
    Process &  $\sqrt{s}$ [GeV] & Stage  & Detector & Performance of \\ \midrule\midrule
    $\PH\PZ, \PZ\to \qqbar$,  & \multirow{2}{*}{350} & \multirow{2}{*}{1}  & \multirow{2}{*}{CLIC\_o3\_v14} & jets at low   energy and low
    \\ ~~$\PH\to$ incl. &&&&  $\gamma \gamma \to$\,had. background\\\midrule

   $\PH \nunubar, \PH\to \mumu$  &
                                                     1400&2
                                         &  CLIC\_ILD
                                                    & muons; dimuon
                                                      mass resolution\\\midrule
    
    \multirow{2}{*}{$    \ww \to \ell \nu \qqbar$}  &\multirow{2}{*}{3000}&\multirow{2}{*}{3}
                                         & \multirow{2}{*}{CLIC\_ILD}
                                                    &jets  and leptons
                                                      at  high
                                                      energy\\  &&&&
                                                                     and high  $\gamma \gamma \to$\,had. background\\
    \midrule

    $\ttbar \PH, \PH\to \bb$, & \multirow{ 2}{*}{1400} & \multirow{
                                                         2}{*}{2}
                                        & \multirow{ 2}{*}{
                                                     CLIC\_SID} & jets
    and leptons with\\
    ~~$\ttbar\to$semileptonic &   & & &    $\gamma \gamma \to$\,had. background   \\\midrule

  \end{tabular}
\end{table}

\subsection{Higgsstrahlung with hadronic Z decay at 350\,GeV}
The process $\epem \to \PZ(\to \qqbar) \PH(\to \text{incl}.) $ is
evaluated at $\sqrt{s}=$350\,GeV. This is an important process for the
CLIC physics programme as the model-independent measurement of Higgs
properties relies on it.
It allows  the performance of the CLICdet card to be validated for jets at
low energy and with low levels of beam-induced \gghadrons background.
For the validation, observables obtained by running \Delphes with the
CLICdet card of the first stage are compared with full detector simulation for CLICdet, version CLIC\_o3\_v14.
In both cases, the same generator level events produced with
\textsc{Whizard} interfaced to \textsc{Pythia6} for the parton shower
and hadronisation are used.
Jets from VLC with $N$\,=\,4, $R$\,=\,1 and $\beta=\gamma=1$ are used unless noted otherwise.

\begin{figure}[hptb]\centering
\includegraphics[width=0.49\textwidth]{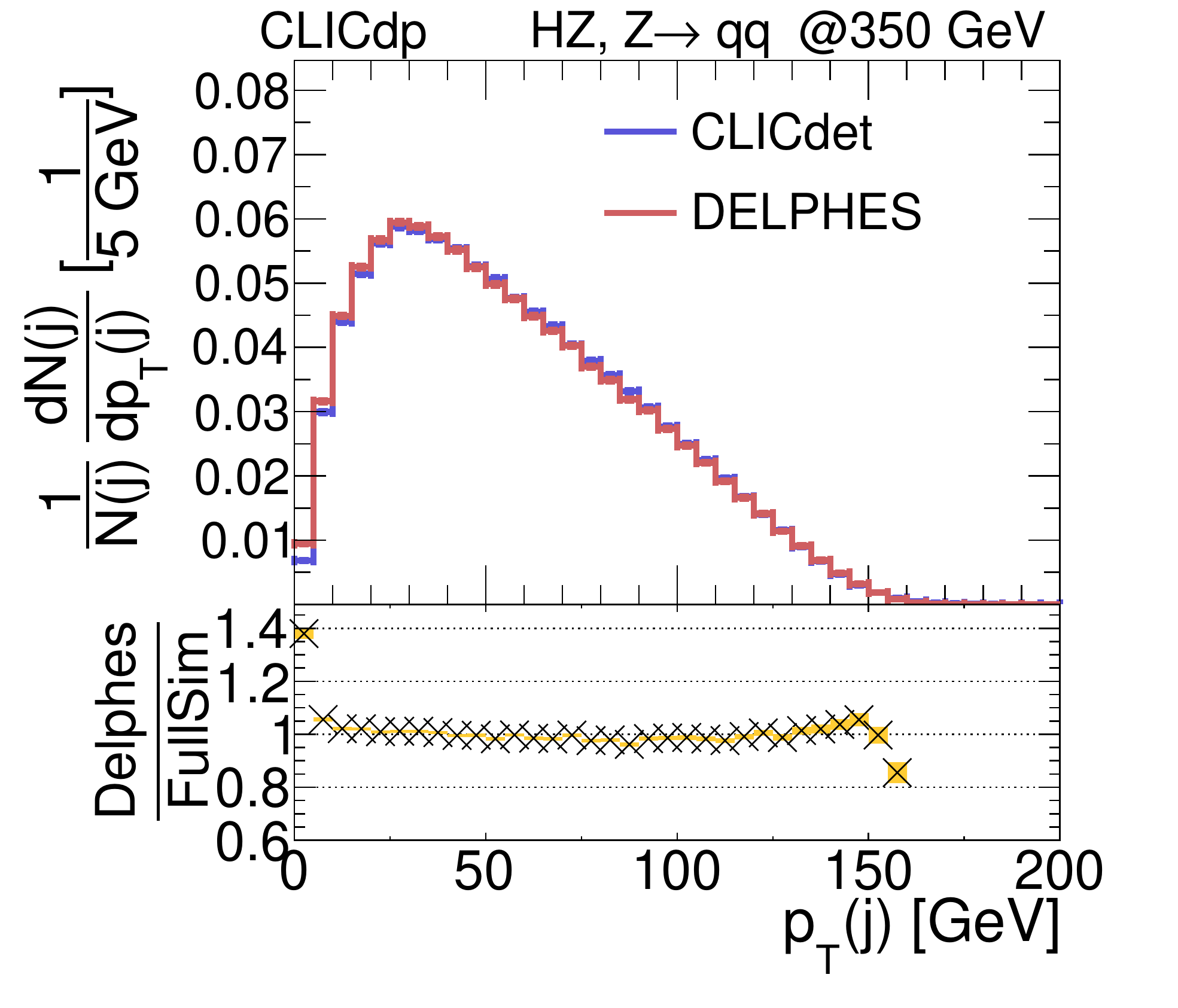}
\includegraphics[width=0.49\textwidth]{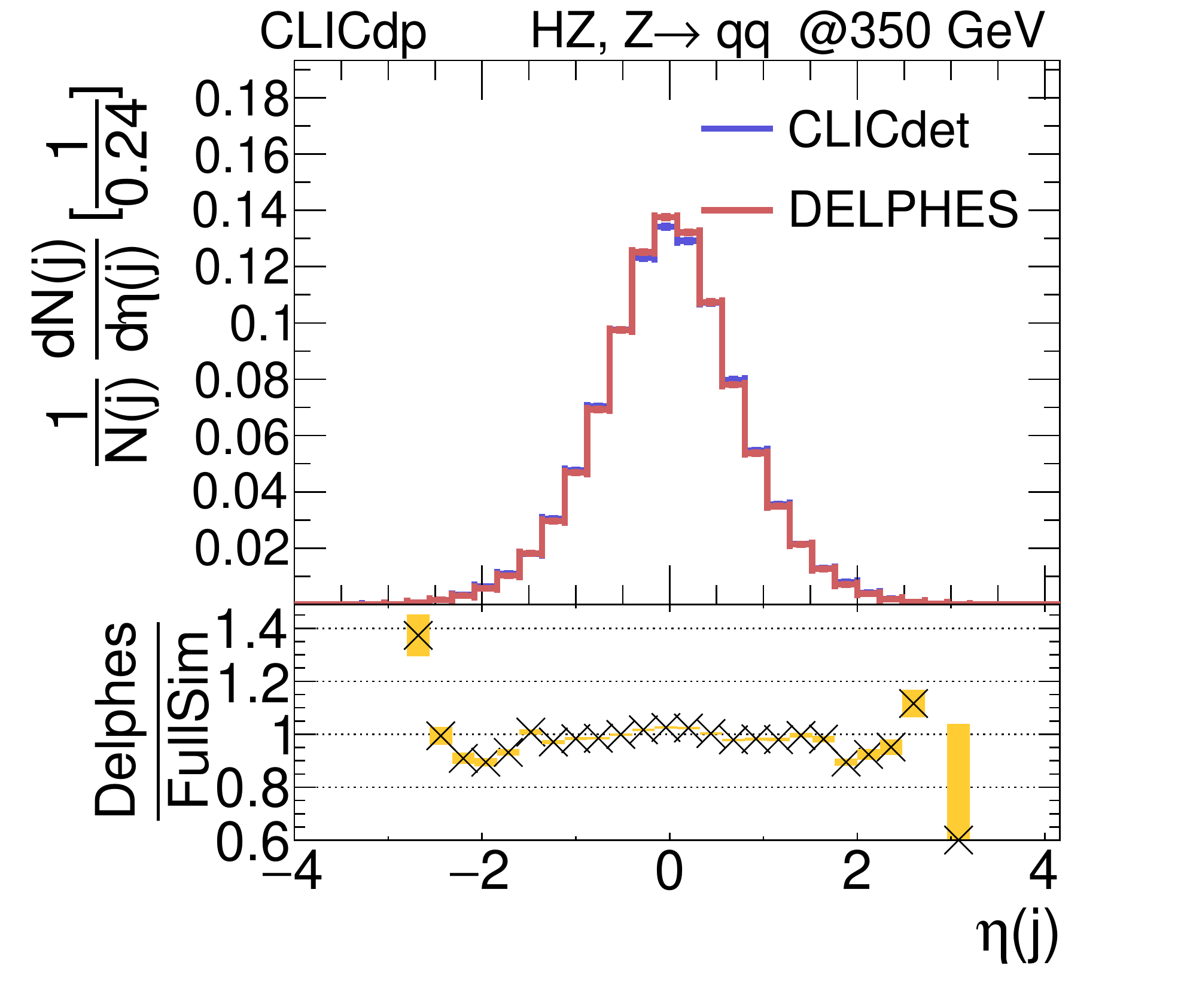}
 \caption[]{Comparison of jet transverse momentum (left) and
   pseudorapidity (right) in Higgsstrahlung events with hadronic Z
   decay for full simulation of CLICdet (blue) and \Delphes (red).}
    \label{fig:HZqqjets}
  \end{figure}

  Jet observables are shown in Fig.~\ref{fig:HZqqjets}.
  Good agreement at the level of better than 5\,\% is found for
  central jets over most of the transverse momentum range above 10\,GeV. Up to
  10\,\% disagreement is found in the forward region $1.5 <
  |\eta|<2.5$. Larger disagreements exist only at low transverse momentum
  $p_{\text{T}}<10$\,GeV and high transverse momentum $p_{\text{T}}> 150$\,GeV.

\begin{figure}[ht]\centering
\includegraphics[width=0.49\textwidth]{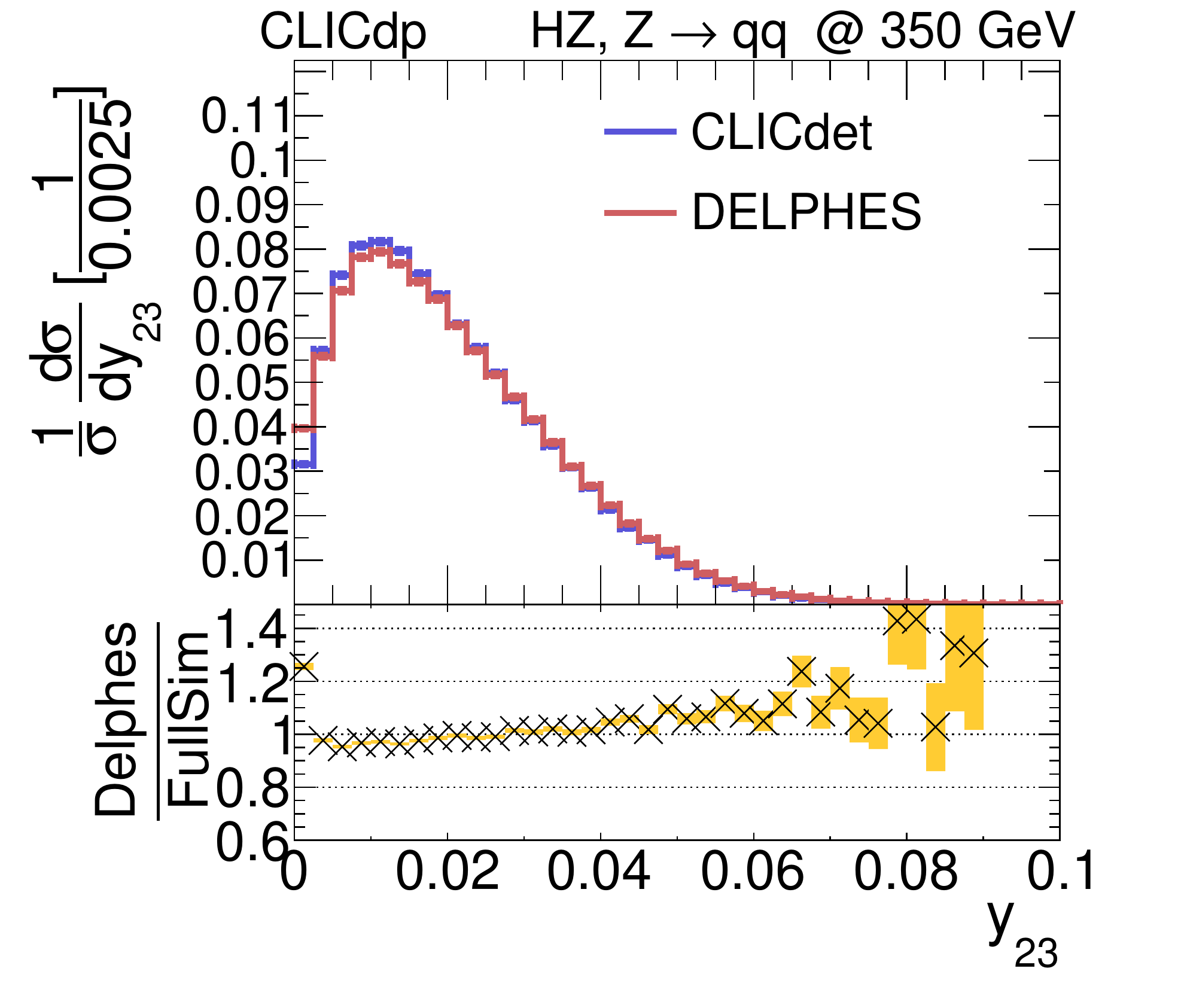}
\includegraphics[width=0.49\textwidth]{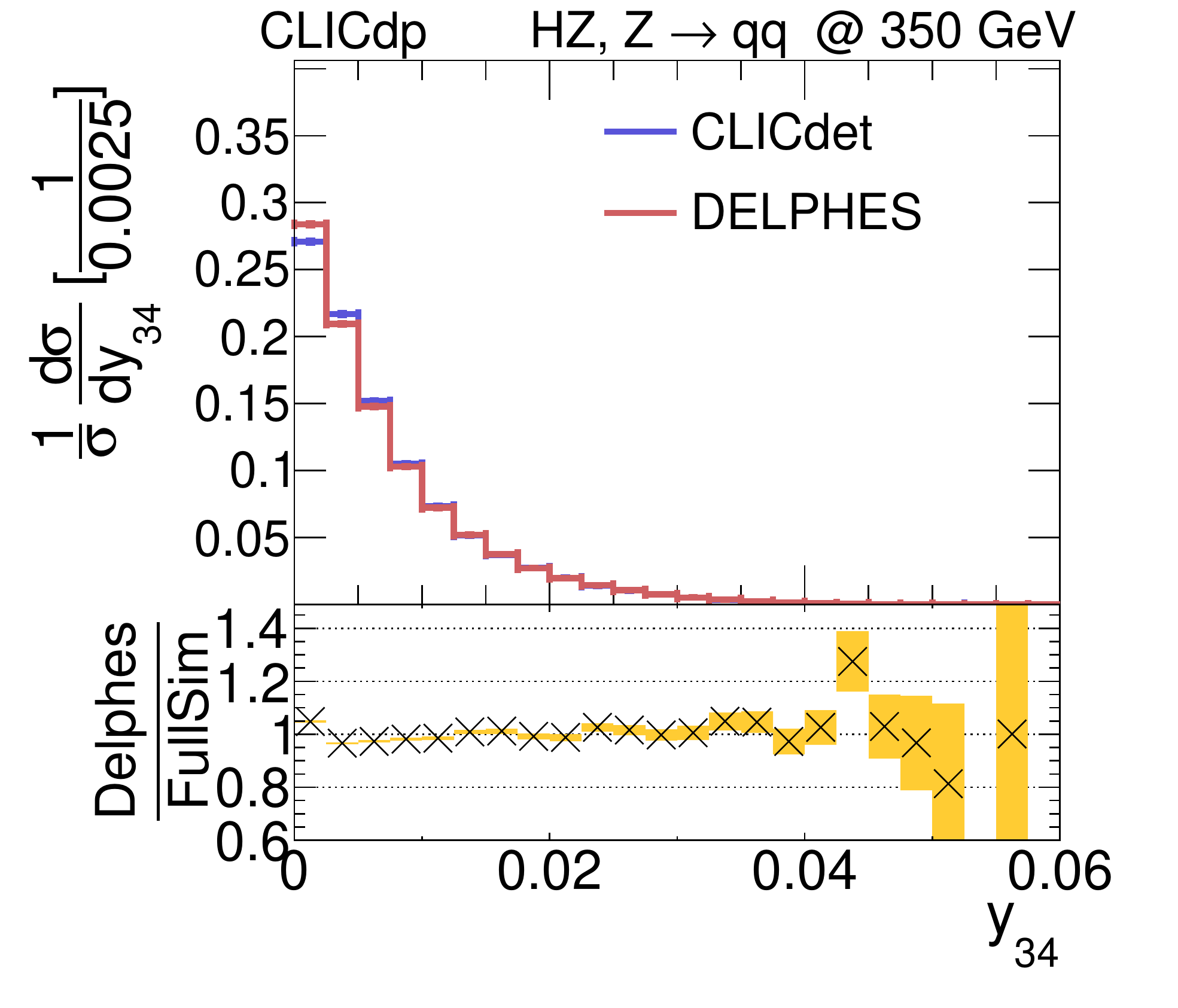}
 \caption[]{Comparison of jet resolution thresholds $y_{23}$ (left) and
   $y_{34}$ (right) in Higgsstrahlung events with hadronic Z
   decay for full simulation of CLICdet (blue) and \Delphes (red).}
    \label{fig:HZqqjetmultipl}
  \end{figure}

Jet resolution threshold observables obtained from the exclusive clustering mode are shown in Fig.~\ref{fig:HZqqjetmultipl}.
The variable $y_{23}$ ($ y_{34}$) is the distance measure associated with
merging from 3 to 2 (4 to 3) jets~\cite[Sec. 3.3.2]{Cacciari:2011ma}.
They are well-modelled apart from a slight
shift to higher values for \Delphes.

\begin{figure}[ht]\centering
\includegraphics[width=0.49\textwidth]{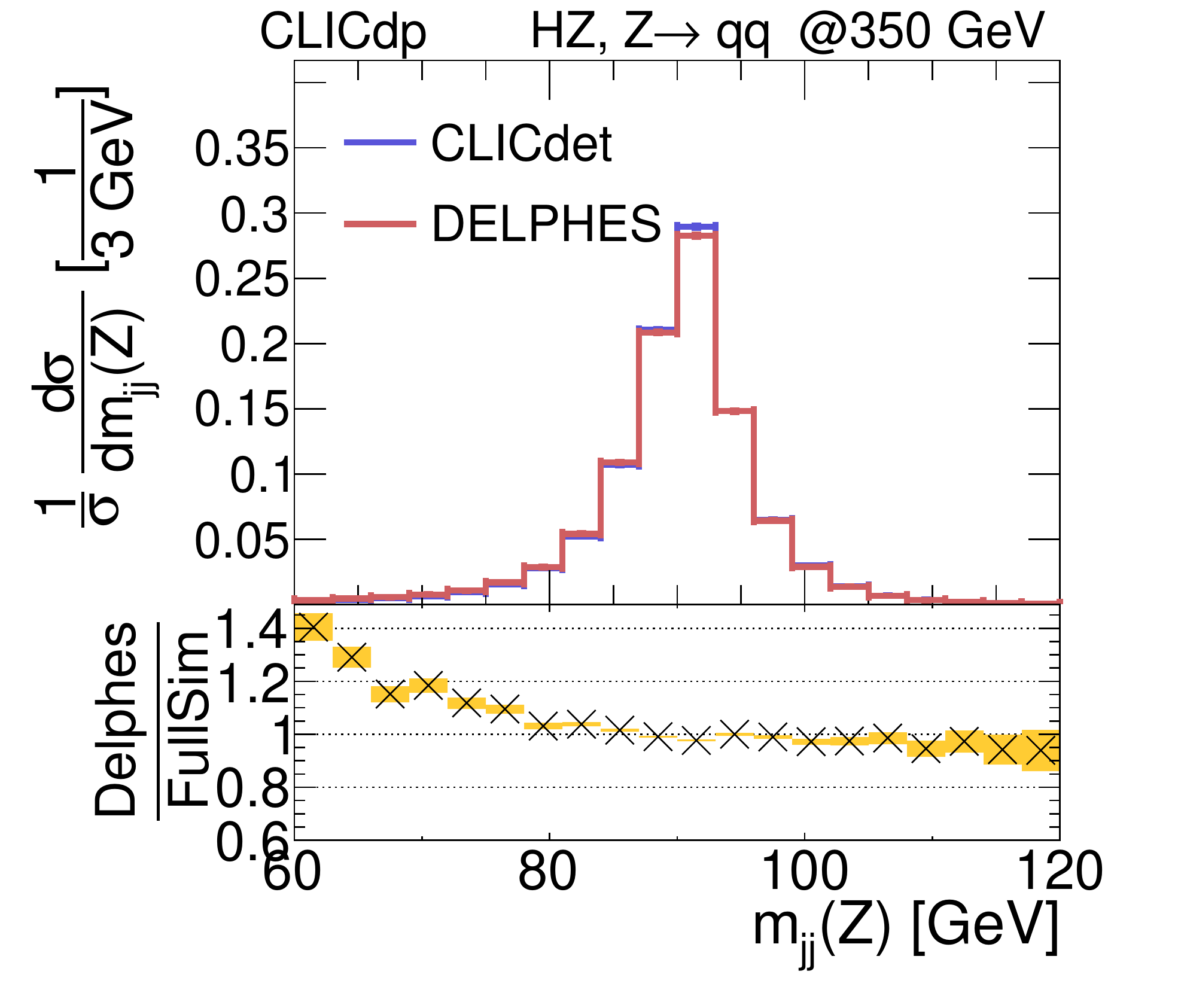}
\includegraphics[width=0.49\textwidth]{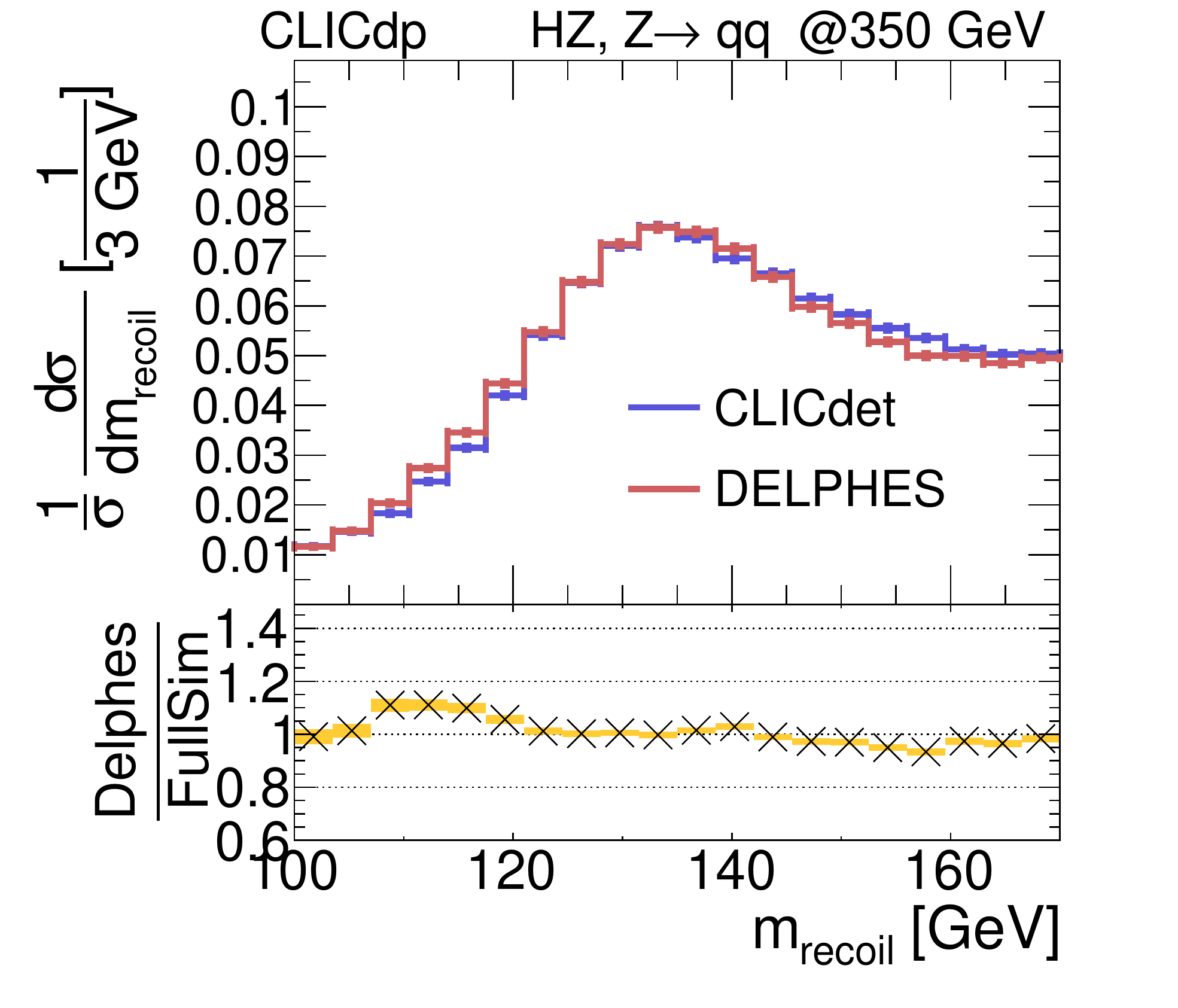}
 \caption[]{Comparison of the di-jet mass of jets assigned to the Z
   boson (left) and the recoil mass (right) in Higgsstrahlung events with hadronic Z
   decay for full simulation of CLICdet (blue) and \Delphes (red).}
    \label{fig:HZqqderived}
  \end{figure}

  The Higgsstrahlung process is further analysed by choosing the
  configuration of two jets with an invariant mass closest to the mass
  of the Z boson, $m_{\PZ}$, among the possible combinations obtained by
  clustering the event with 2, 3, and 5 jets exclusively.
  Fig.~\ref{fig:HZqqderived} shows the good agreement of the di-jet
  invariant mass for Z jets (left) and the recoil mass (right)
  between full simulation and \Delphes.
   Systematic differences mostly below 10\,\% are seen in the low tail of the invariant
   mass distribution and in the slopes of the recoil mass.
Additional kinematic distributions can be found in Appendix~\ref{app:extraplots_hz}.

  \subsection{Semi-leptonic top-quark pairs associated with a Higgs boson at 1.4\,TeV}
  The process $\epem \to \ttbar \PH $  with semi-leptonic decay of the
  $\ttbar$ pair and Higgs decays to $\bb$ at 1.4\,TeV collision energy is simulated with the
  CLIC\_SiD detector model~\cite{cdrvol2}, including the beam-induced
  background from \gghadrons.
  Jet clustering is performed with the VLC algorithm with $R\,=\,1.0$
  and $N$\,=\,6.
  This process allows validation of the performance of leptons and jets,
  including the effect of beam-induced background.

  \begin{figure}[ht]\centering
  \includegraphics[width=0.495\textwidth]{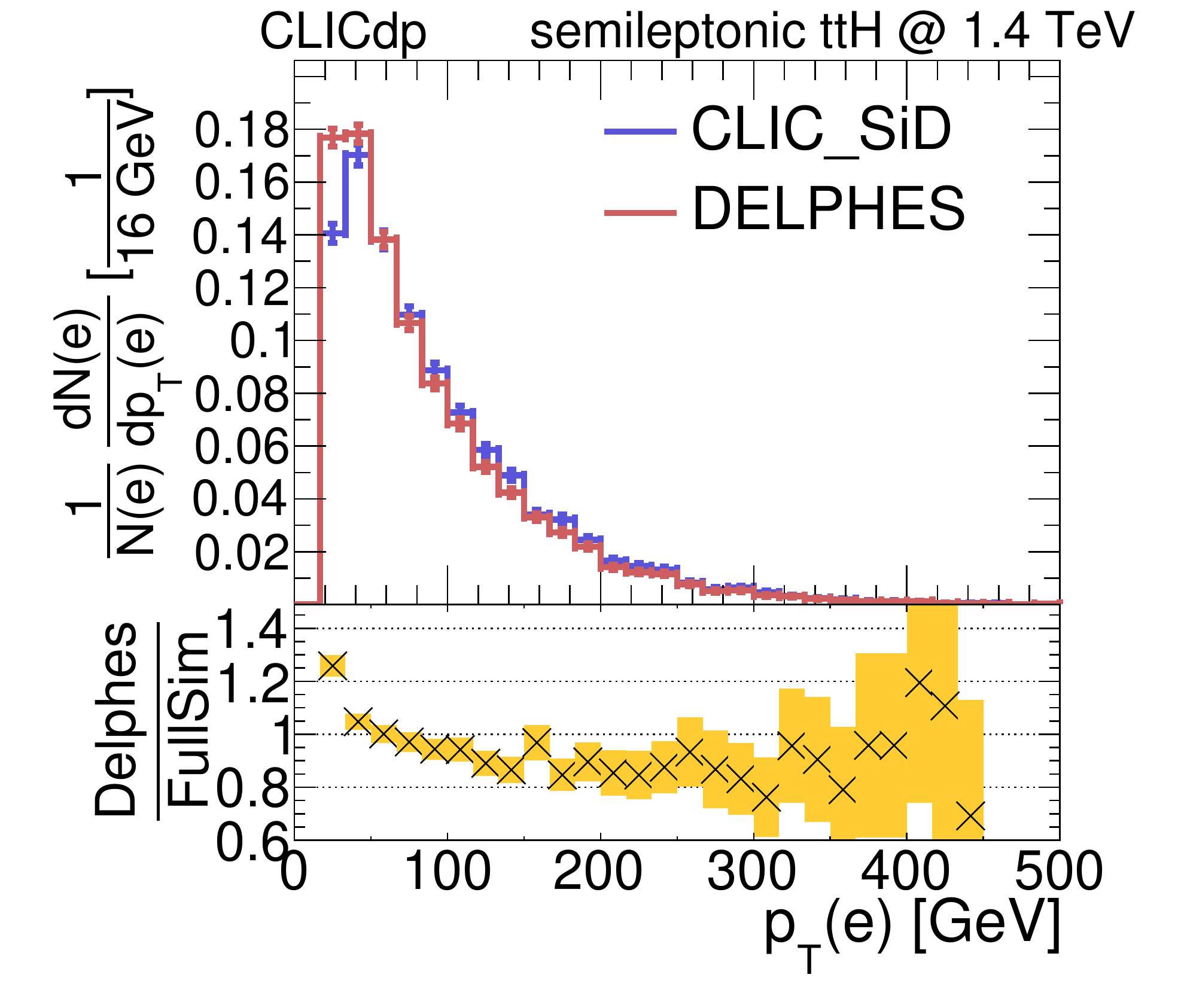}
      \includegraphics[width=0.495\textwidth]{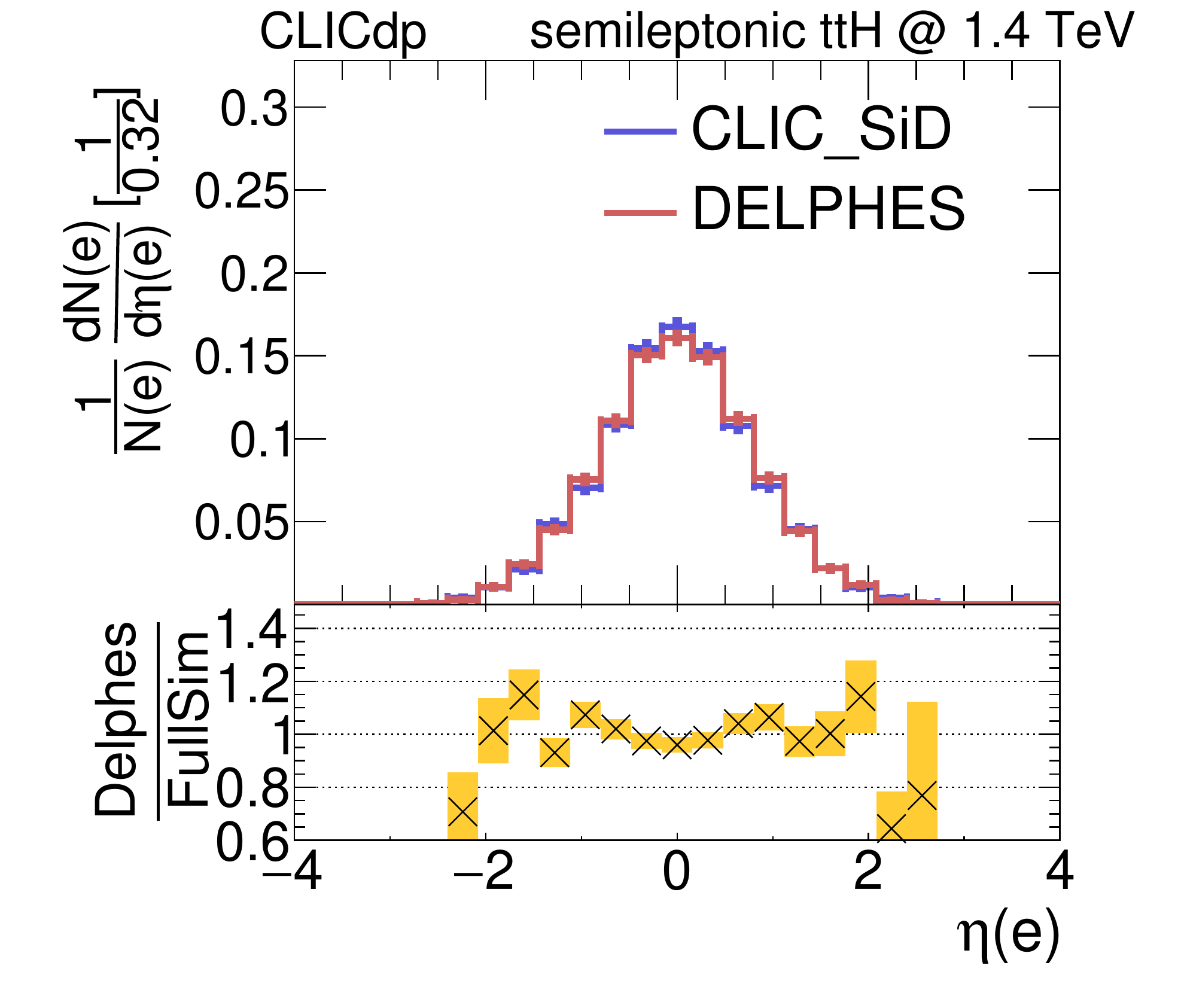}
      \caption[validation of electron performance in ttH]{Comparison of
        electron transverse momentum (left) and pseudorapidity (right) in semileptonic $\ttbar \PH$ events for full
        simulation of CLIC\_SiD (blue) and \Delphes (red).}
      \label{fig:tthelectrons}
    \end{figure}

  In the full simulation analysis, optimised isolation criteria are used
  which give good performance down to low momenta.
  A cut on the transverse momentum of electrons and muons requiring at
  least 20\,GeV selects the region where the cone-based isolation with
  a transverse momentum ratio cut as implemented in \Delphes is
  applicable.
  The performance of the CLICdet \Delphes card for electrons is
  demonstrated in Fig.~\ref{fig:tthelectrons}.  
  Good agreement can be seen in the transverse momentum of the electron,
  Fig.~\ref{fig:tthelectrons}\,(left), with the \Delphes distribution
  shifted to smaller values of transverse momentum.
  Fig.~\ref{fig:tthelectrons}\,(right) shows the pseudorapidity
  distribution, which is well-modelled. 
 The performance of the fast simulation for muons and jets is
 presented in Appendix~\ref{app:extraplots_ttH}. 
   
\subsection{Di-boson production WW at 3\,TeV}
\begin{figure}[ht]\centering
  \includegraphics[width=0.49\textwidth]{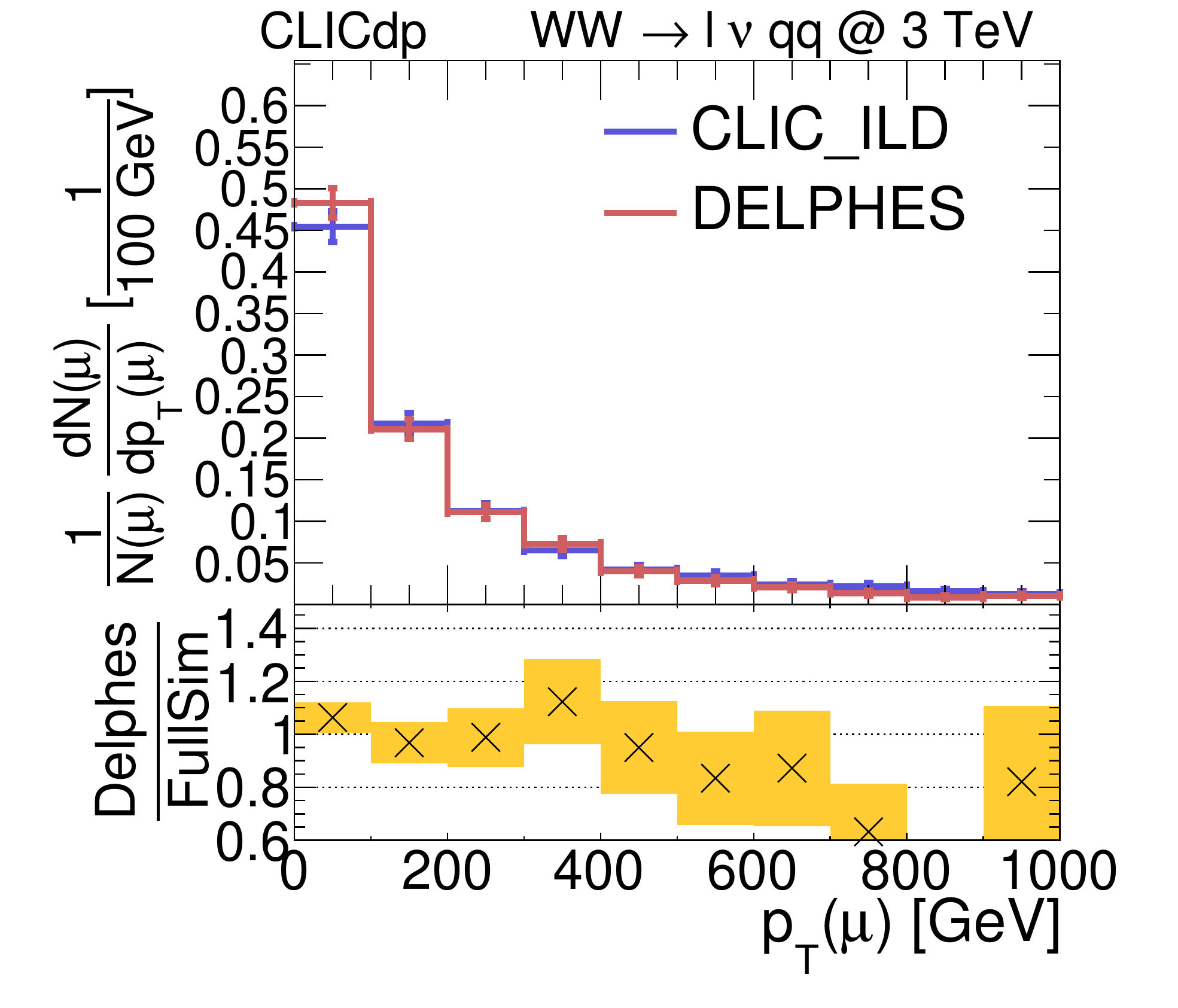}
  \includegraphics[width=0.49\textwidth]{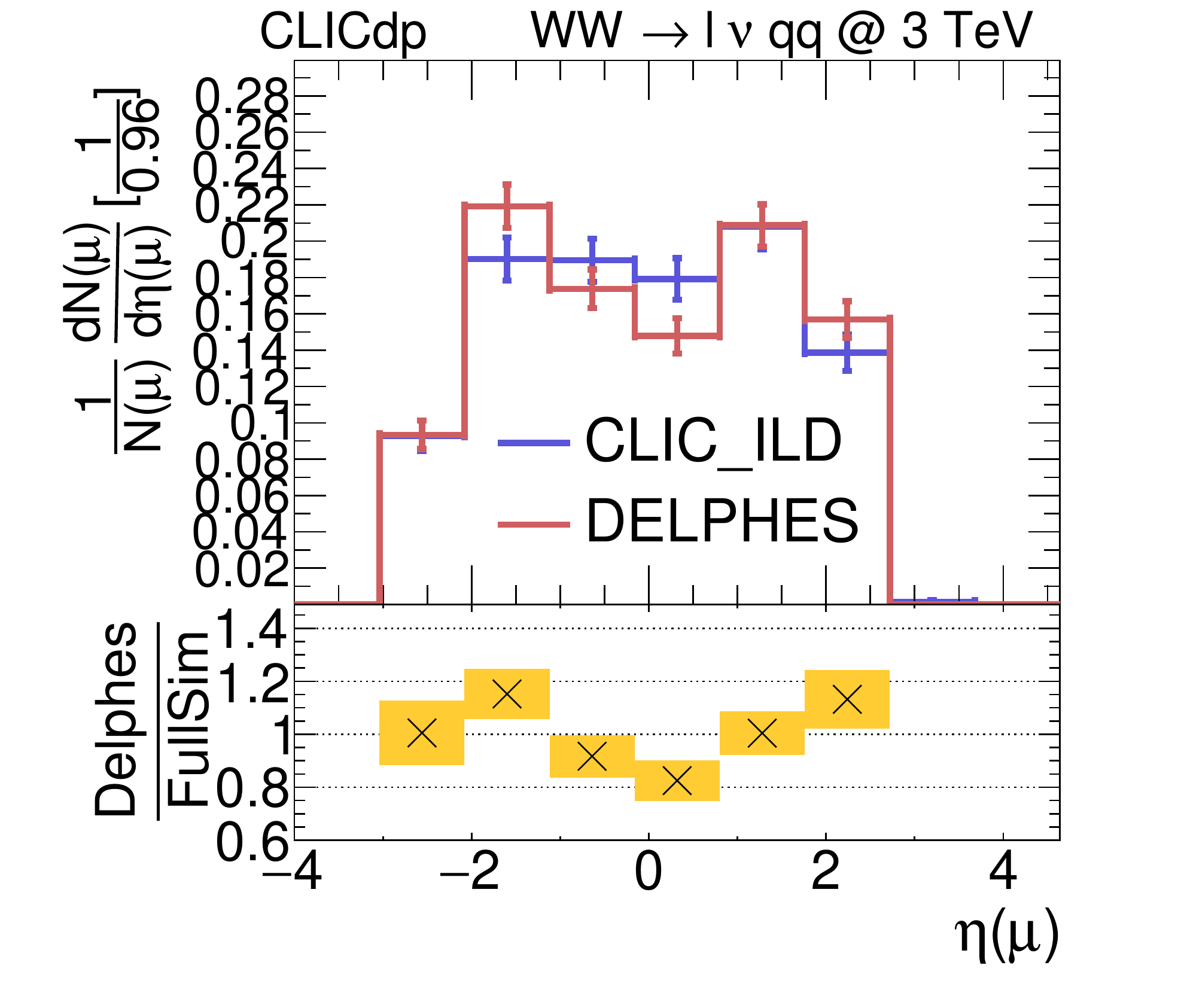}
  \caption[]{Validation of fast simulation performance for the muon
    transverse momentum (left) and pseudorapidity (right) for events with the final state $\ell \nu \qqbar$ at 3\,TeV
    with full simulation of CLIC\_ILD (blue) and fast simulation
    with \Delphes (red).}
  \label{fig:WWmuons}
\end{figure}

The performance of fast simulation for muons and jets is validated
using the process $\epem \to \ell \nu \qqbar $ at 3\,TeV.
Jet clustering is performed with the VLC algorithm using $R$\,=\,0.7 and
$N$\,=\,2. Beam induced \gghadrons background is overlayed in the full
    simulation, and its effects are taken into account in the fast simulation.
Figure~\ref{fig:WWmuons} demonstrates the validation of the muon
simulation.
The transverse momentum  and
pseudorapidity are in agreement
between full and fast simulation.

\begin{figure}[hpbt]\centering
    \includegraphics[width=0.49\textwidth]{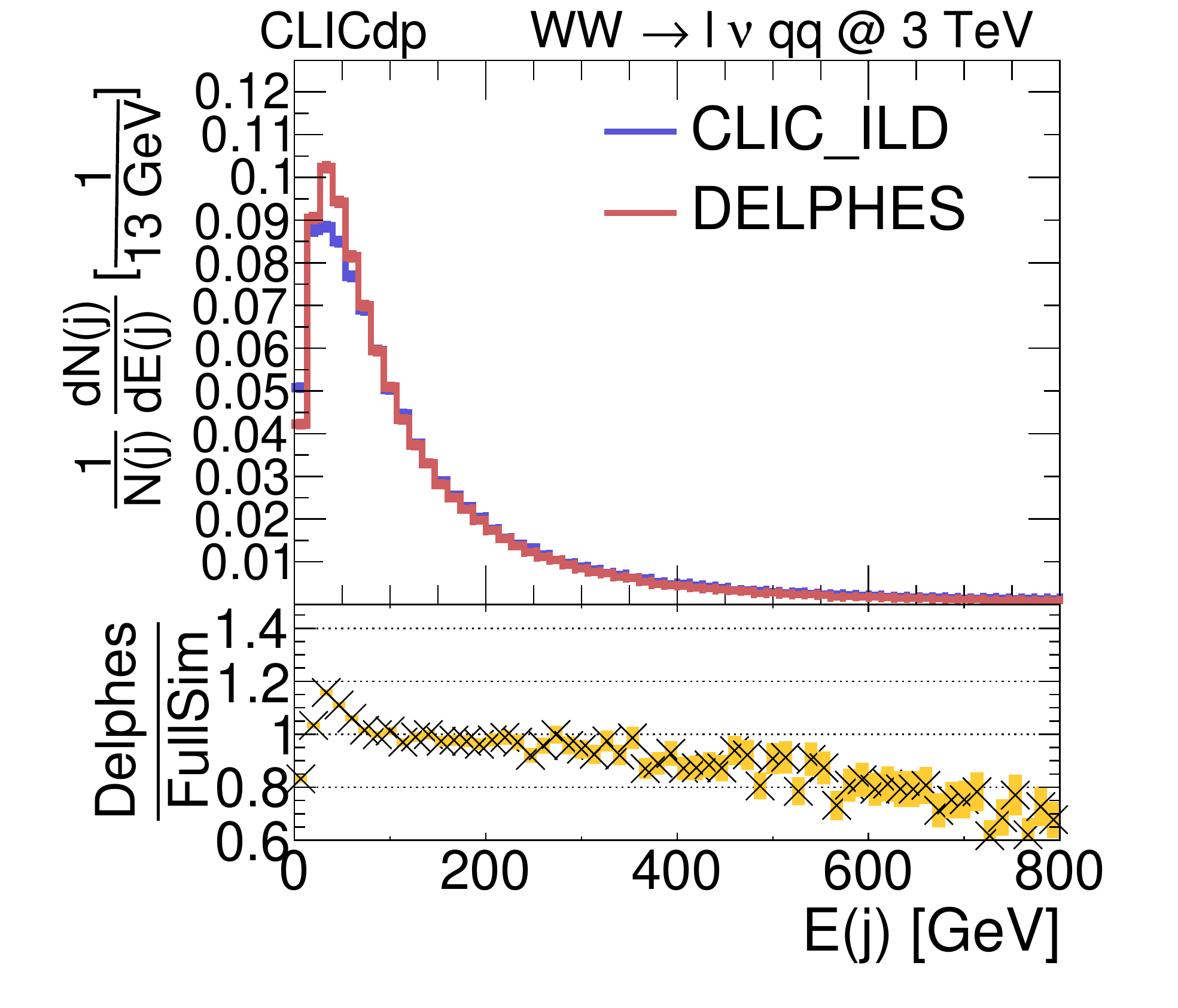}
   \includegraphics[width=0.49\textwidth]{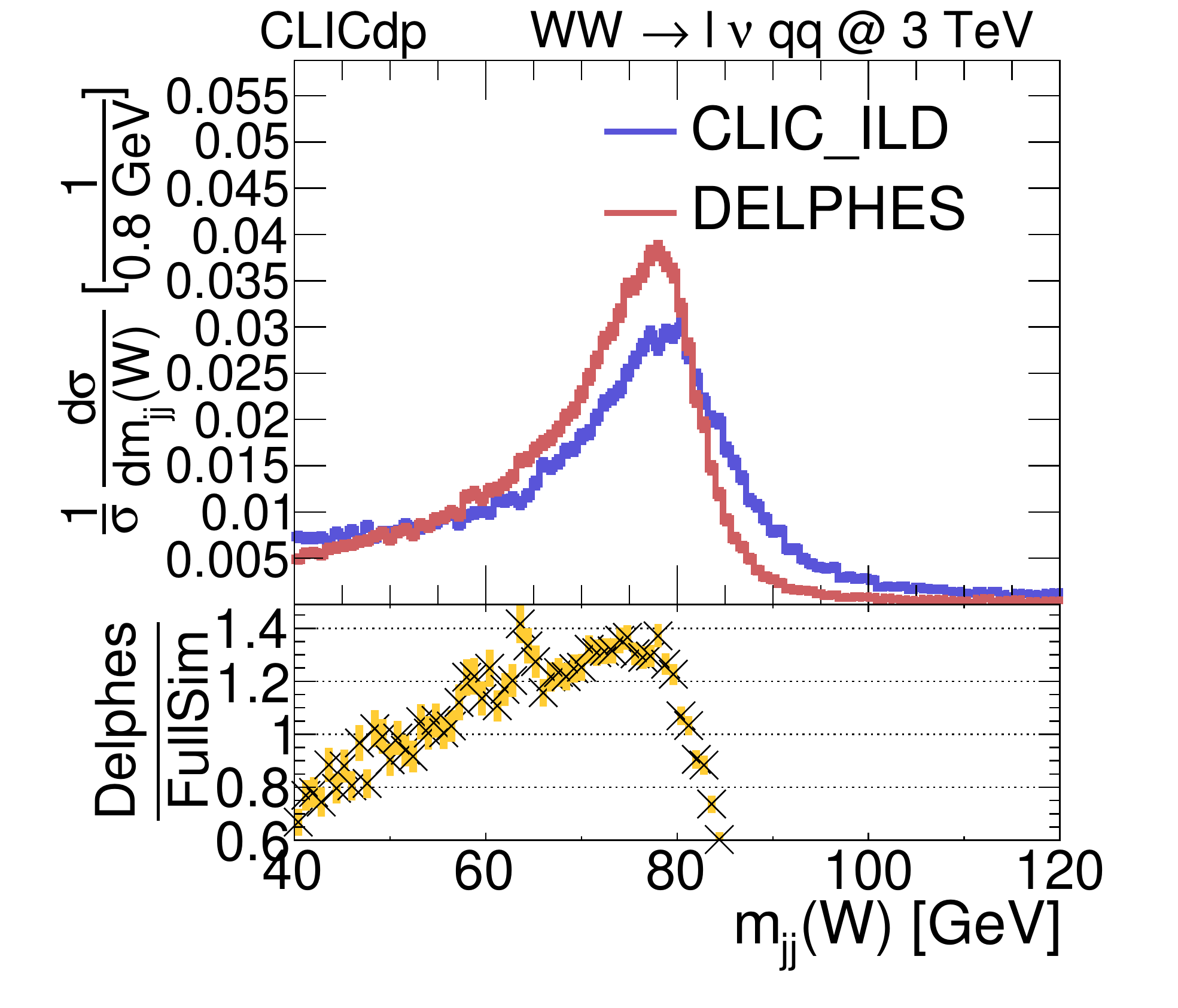}
    \caption[]{Validation of fast simulation performance for jet energy
      (left) and  di-jet invariant mass (right)  using  events with the final state $\ell \nu \qqbar$ at 3\,TeV
      with full simulation of CLIC\_ILD (blue) and fast simulation
      with \Delphes (red).}
    \label{fig:WWjets}
  \end{figure}

  The performance of the \Delphes simulation for jets at CLIC is
  presented in Fig~\ref{fig:WWjets}.
The energy is shifted to lower values for \Delphes.
The invariant mass of the two jets, associated with
the $\PW$ boson, is reconstructed with a narrower width and shifted to
lower values in \Delphes compared to the full simulation.
As the jet clusters all of the typically soft particles from the background distributed
within its entire area, the effect on the mass due to their angular
spread is larger than the effect on
the energy.
The description could be improved by smearing not only the energy but
also the mass.
Alternatively, one could overlay \gghadrons events.
However, highly granular timing information from the detector is crucial to suppress
contributions from \gghadrons.
The current implementation cannot estimate this suppression and its
impact on the reconstruction of the signal process
since no timing information is available in the \Delphes simulation.

Additional kinematic distributions of jets can be found in Appendix~\ref{app:extraplots_ww}.

\subsection{Di-muon mass resolution in Higgs decays to muon pairs at 350\,GeV and 1.4\,TeV}

\begin{figure}[ht]\centering
  \includegraphics[width=0.49\textwidth]{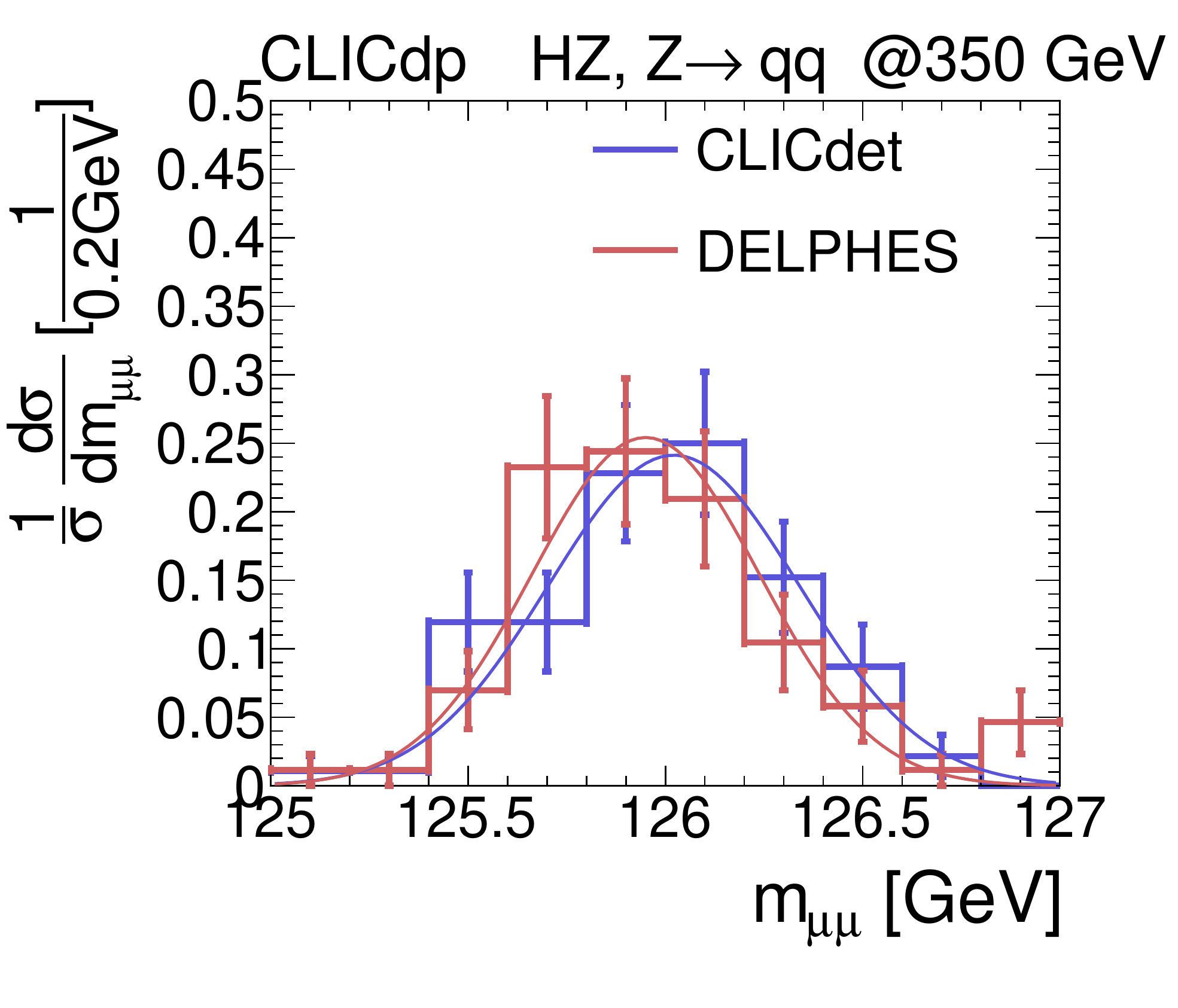}
  \includegraphics[width=0.49\textwidth]{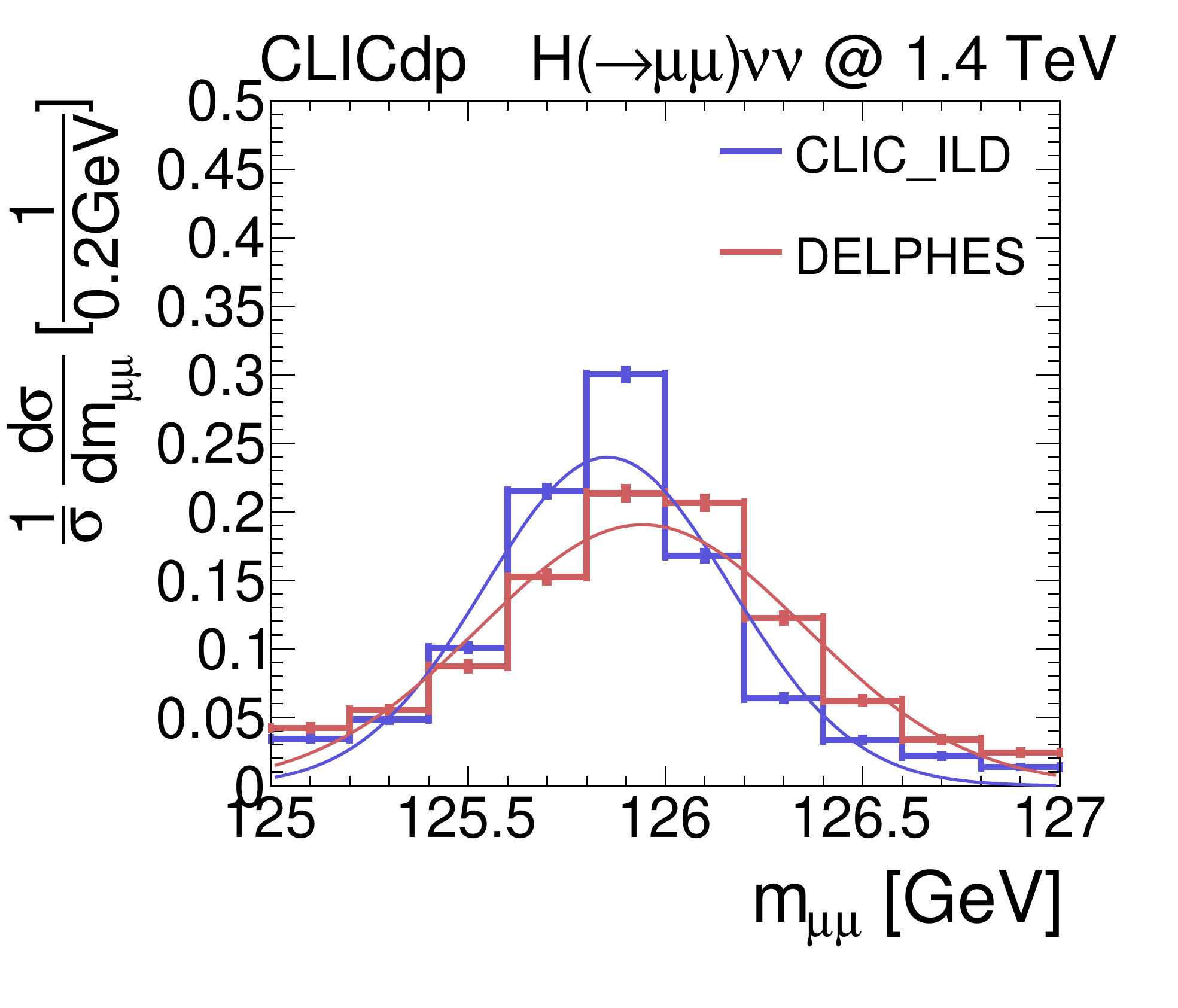}
  \caption[]{Validation of the di-muon peak obtained in $H \to \upmu
    \upmu$ decays at in $\PH\PZ$ events at 350\,GeV (left) and in $\PH \nu
    \nu$ events at 1.4\,TeV (right). Full detector simulation events (blue) are compared with events produced by applying the CLICdet \Delphes card (red). }
    \label{fig:dimuonmass}
  \end{figure}

The di-muon mass resolution is validated in  $H \to \upmu \upmu$ decays at 350\,GeV and 1.4\,TeV.
The Higgs boson resonance peaks in the di-muon mass distributions for
the two energy stages are shown in Fig.~\ref{fig:dimuonmass}. The corresponding mean and width obtained with a Gaussian fit are presented in Table~\ref{tab:dimuonfit}.
The resulting  mean and width of the Gaussian di-muon peak are in good
agreement for $\PH(\to \upmu \upmu) \PZ(\to \qqbar)$ events at 350\,GeV and in
agreement within 30\,\% for  $\PH (\to \mu \mu) \nu \nu$ events at
1.4\,TeV.
As shown in~\cite[Fig.\,32]{DetectorPerformance2018}, tracking
resolutions in the forward region are less well described by the assumed
parametrisation
than in the central region.
As the muons in $\PW$-boson fusion $\PH (\to \upmu \upmu) \nu \nu$ events are more forward in
the detector than those in $\PH(\to \upmu \upmu) \PZ(\to \qqbar)$
events,
this explains the larger discrepancy between \Delphes and full
simulation for the di-muon mass in the $\PW$-boson fusion $\PH (\to \mu \mu) \nu \nu$ process.
\begin{table}[htbp]
  \centering
  \caption{Resulting parameters of a Gaussian fit to the invariant di-muon
    mass in $\PH \to \upmu \upmu $ events}
  \label{tab:dimuonfit}
  \begin{tabular}[ht]{lcccccc}
\toprule
    Simulation & \multicolumn{2}{c}{$\PH(\to \upmu \upmu) \PZ(\to \qqbar)$ at 350 GeV} & \multicolumn{2}{c}{$\PH (\to \mu \mu) \nu \nu$ at 1.4 TeV}\\ \midrule
       & $\mu$ [GeV]& $\sigma$ [GeV]& $\mu$ [GeV] & $\sigma$ [GeV]\\ \midrule
      Full detector & $126.0\pm 0.03$ & $0.32\pm 0.03$ &$125.8\pm 0.003$ &$0.31 \pm 0.003$\\
      \Delphes & $125.9\pm0.04$  &  $0.29 \pm 0.03$& $125.9 \pm 0.004$ & $0.41 \pm 0.005$\\\bottomrule
  \end{tabular}
\end{table}

\section{Conclusions}
Three parameter cards for the detector model for CLIC, CLICdet, have been prepared for the \Delphes framework.
They describe the CLICdet geometry and performance for the three energy stages of CLIC.
Performance parameters are based on full simulation studies of the detector response and reconstruction software.
The cards have been validated by comparing with full simulation for various processes.
At the first energy stage of CLIC with low levels of beam-induced background, the validation has found  good agreement in particular for single object observables. Also derived observables such as di-jet invariant masses are well described in the regions. At the higher energy stages, reasonable agreement has been found for transverse momentum, energy, and pseudorapidity distributions of leptons and jets. The absence of the beam-induced background in the fast simulation limits the ability to reproduce all observables fully accurately. A smearing of the jet energy has been applied to mimic the effects of \gghadrons background. This could be improved in the future by applying a smearing also to the jet mass.

As \Delphes has so far been used mainly for studies at hadron colliders, some features of detectors at lepton colliders had not yet been implemented.
Exclusive jet clustering and the VLC algorithm have been added as part of this study.
In addition, implementations for $c$-tagging and for the very forward calorimeters as well as a refinement of the isolation criteria could be added in the future.
The possibility of overlaying beam-induced \gghadrons events would improve  the description of  detectors at linear colliders.
However, this would also require to add the simulation of time information to exploit the timing selection capabilities of the detector to suppress these background contributions.

Full simulation studies for future experiments remain indispensable for the evaluation of the detector performance.
Furthermore, they are necessary for analyses using observables that are sensitive to beam-induced backgrounds and rely on timing information, analyses based on tails of distributions that are not well-modelled by fast simulation,  as well as analyses using exotic signatures such as long-lived particles.

In conclusion, a good description of the detector concept for CLIC has been achieved in the \Delphes fast simulation framework with the parameter cards discussed in this note.
These cards have been validated and have already been used for CLIC physics studies.
The implementation makes it possible to take advantage of the flexibility and accessibility of \Delphes for studying many aspects of the CLIC physics potential.

\section*{Acknowledgements}
We thank  Michele Selvaggi for helpful discussions.
This work benefited from services provided by the ILC Virtual Organisation, supported by the national resource providers of the EGI Federation. This research was done using resources provided by the Open Science Grid, which is supported by the National Science Foundation and the U.S. Department of Energy's Office of Science.

\printbibliography[title=References]  
\newpage
\appendix
\section{Instructions for using the CLICdet \Delphes cards}\label{sec:instructions}
There are three cards for the three energy stages of CLIC. The user
chooses the one appropriate for the relevant energy stage as indicated in Table~\ref{tab:stages}.
    \begin{table}[ht]
      \centering
      \caption{CLIC energy stages implemented in separate CLICdet
        DELPHES cards.}
      \begin{tabular}[ht]{lrl}\toprule
        Stage &	Energy options& Card\\\midrule
        Stage1& 	380 (350) GeV& \texttt{delphes\_card\_CLICdet\_Stage1.tcl}\\
        Stage2& 	1.5 (1.4) TeV& \texttt{delphes\_card\_CLICdet\_Stage2.tcl}\\
        Stage3& 	3 TeV& \texttt{delphes\_card\_CLICdet\_Stage3.tcl}\\
\bottomrule
      \end{tabular}\label{tab:stages}
    \end{table}
To understand the output of running the card, the $b$-tagging and the correct  choice of jet branches is described below.

\subsection{B-Tagging in the CLICdet DELPHES card}
   B-tagging is implemented in the card with three working points
   (WP), which tag 50\%, 70\%, and 90\% of all b-quark-initiated jets,
   respectively. The corresponding mis-tagging rates are implemented
   in the CLICdet card. \Delphes uses a three-bit mask to store the b-tagging information in the following way, where
   \begin{itemize}
   \item bit 0 is the tight WP with 50\% b-tagging efficiency, \item bit 1 is
     the medium WP with 70\% b-tagging efficiency,\item bit 2 is the loose WP
     with 90\% b-tagging efficiency.
\end{itemize}
This leads to the combinations indicated in Table~\ref{tab:bitsforbtag}.
\begin{table}[htbp]
  \centering
  \caption{B-tagging working points in bit-wise code with T for true
    and F for false.}
  \label{tab:bitsforbtag}
   \begin{tabular}[ht]{|l|c|c|c|c|c|c|c|c|}\hline
BTag value      &	7 &  	6 	&5 	&4 	&3 	&2 	&1 	&0   \\\hline
BTag binary     &	111& 	110 	&101 	&100 	&011 	&010 	&001 	&000\\\hline
(bit 0) \& BTag &	T &	F 	&T 	&F 	&T 	&F 	&T 	&F\\
(bit 1) \& BTag &	T &	T 	&F 	&F 	&T 	&T 	&F 	&F\\
(bit 2) \& BTag &	T &	T 	&T 	&T 	&F 	&F 	&F 	&F\\\hline
   \end{tabular}
\end{table}
   Loosely b-tagged jets are selected with \texttt{BTag}$\ge$4; for
   medium b-tagged jets, \texttt{BTag} should be 2, 3, 6, or 7, and for the tight WP the \texttt{BTag} should be 1, 3, 5 or 7.

\subsection{Taking into account the effect of beam-induced background}
 The levels of beam-induced \gghadrons events are higher at the high energy stages.
 To account for their effects,
 a jet energy smearing is applied in the two cards for Stage2 and Stage3.

 The branches with the jet energy smearing applied are then named \textbf{\texttt{JER\_VLCjetR(r)N(njets)}} in the resulting root file.

\section{Supplementary validation studies}\label{app:extraplots}
\subsection{Di-jet mass in Higgsstrahlung with hadronic Z decay at 350\,GeV}\label{app:extraplots_hz}

\begin{figure}[ht]\centering
  \includegraphics[width=0.49\textwidth]{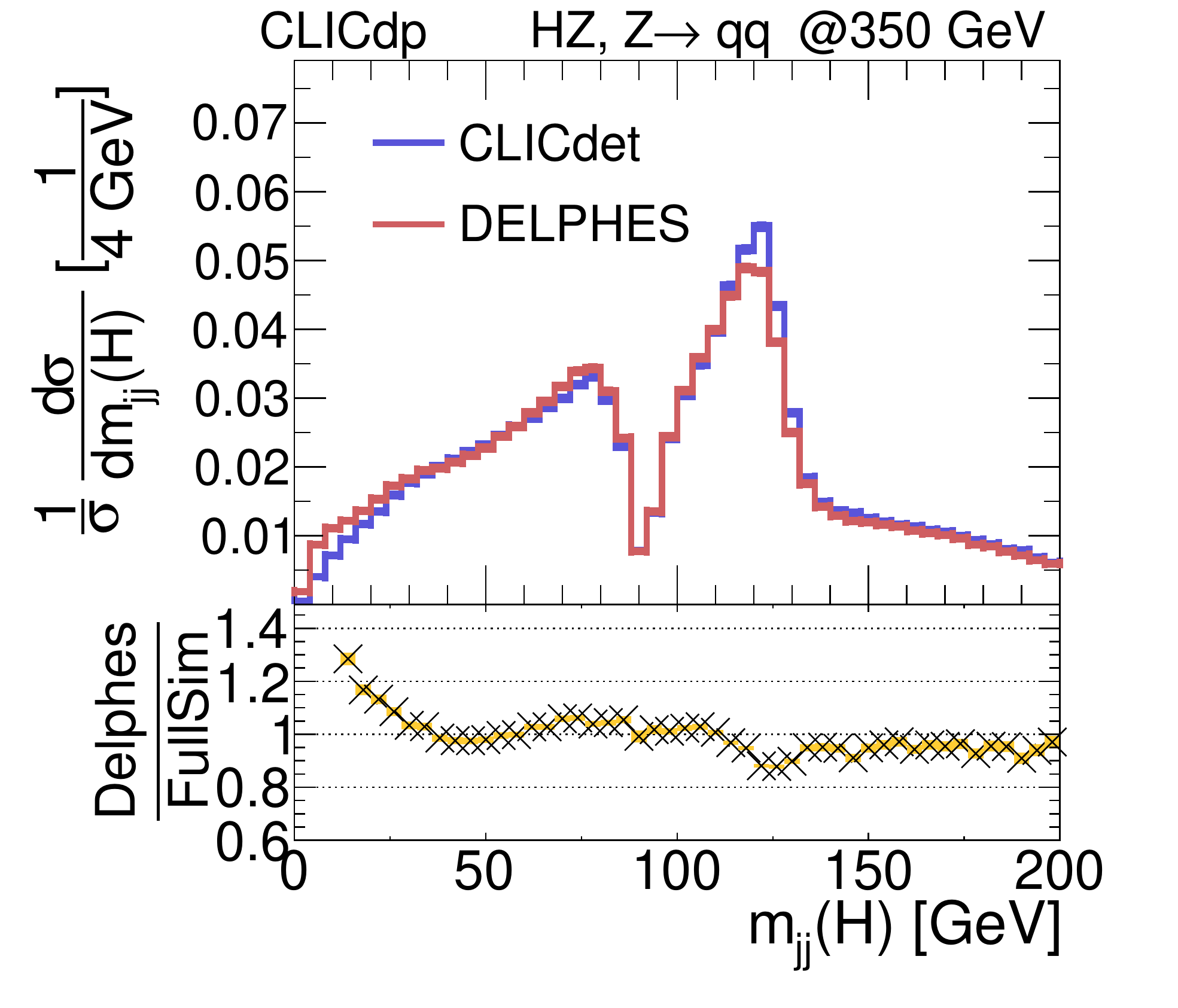}
  \caption[]{Comparison of the di-jet mass  of
   remaining jets not assigned to the Z boson  in Higgsstrahlung events with hadronic Z
   decay for full simulation of CLICdet (blue) and \Delphes (red).}
    \label{fig:HZqqderived_app}
  \end{figure}

Using exclusive jet clustering with N\,=\,4, the Z~jets are assigned
according to the combination closest to the mass of the Z boson.
The invariant mass of
  the two remaining jets after assigning the Z jets is shown in
  Fig.~\ref{fig:HZqqderived_app}(right) as $m_{\text{jj}}(\PH)$.

  \subsection{Semi-leptonic top-quark pairs associated with a Higgs boson at 1.4\,TeV}\label{app:extraplots_ttH}

\begin{figure}[ht]\centering
  \includegraphics[width=0.495\textwidth]{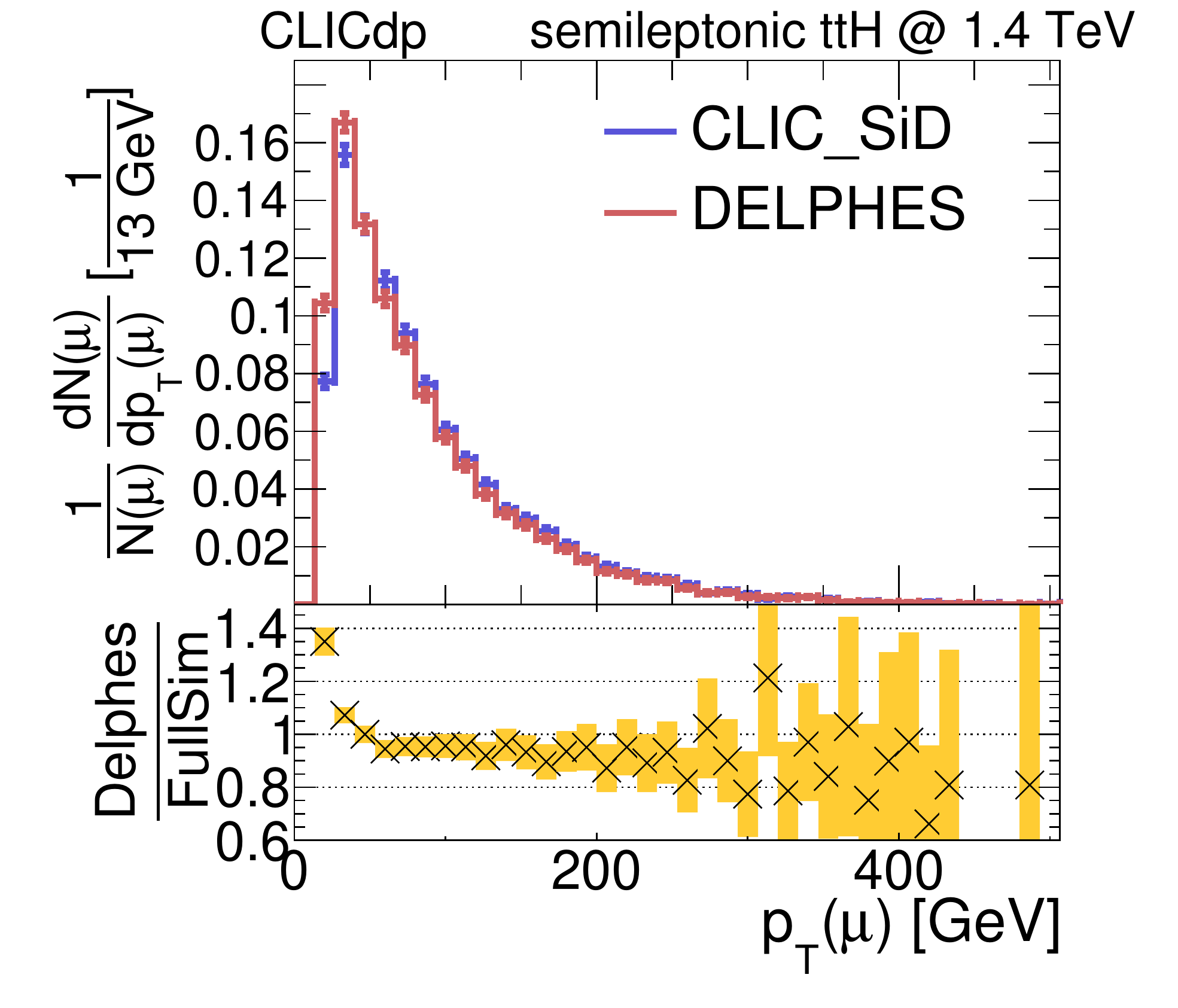}
      \includegraphics[width=0.495\textwidth]{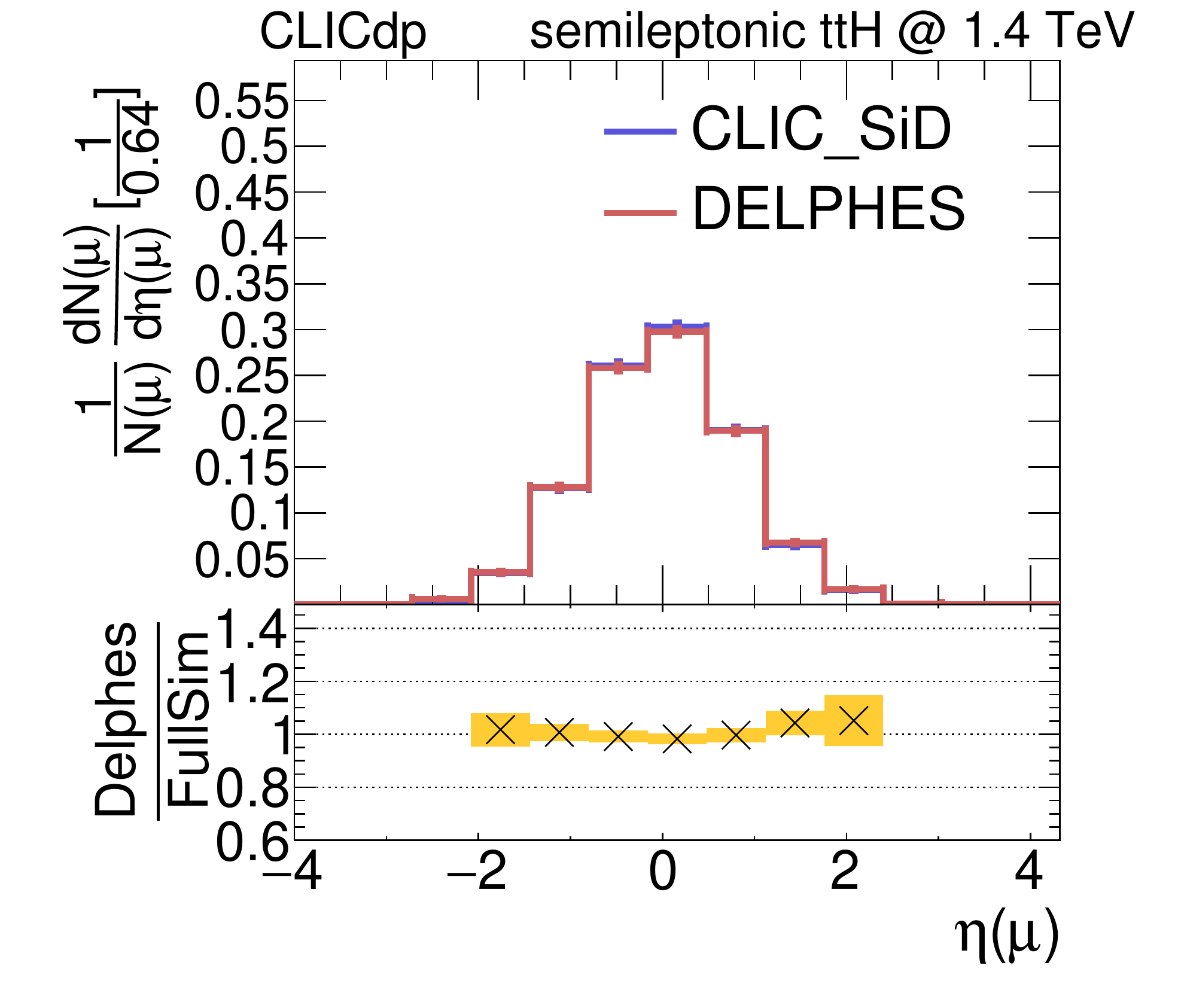}
      \caption[]{Comparison of
    muon transverse momentum (left) and pseudorapidity (right) in semileptonic $\ttbar \PH$ events for full
    simulation of CLIC\_SiD (blue) and \Delphes (red).}
      \label{fig:tthmuons_app}
    \end{figure}

The fast simulation for muons is validated in
Fig.~\ref{fig:tthmuons_app}. Like for electrons
(Fig.~\ref{fig:tthelectrons}),
the transverse momentum is described reasonably well, but
with an increase in contributions from \Delphes at lower values.

\begin{figure}[ht]\centering
    \includegraphics[width=0.495\textwidth]{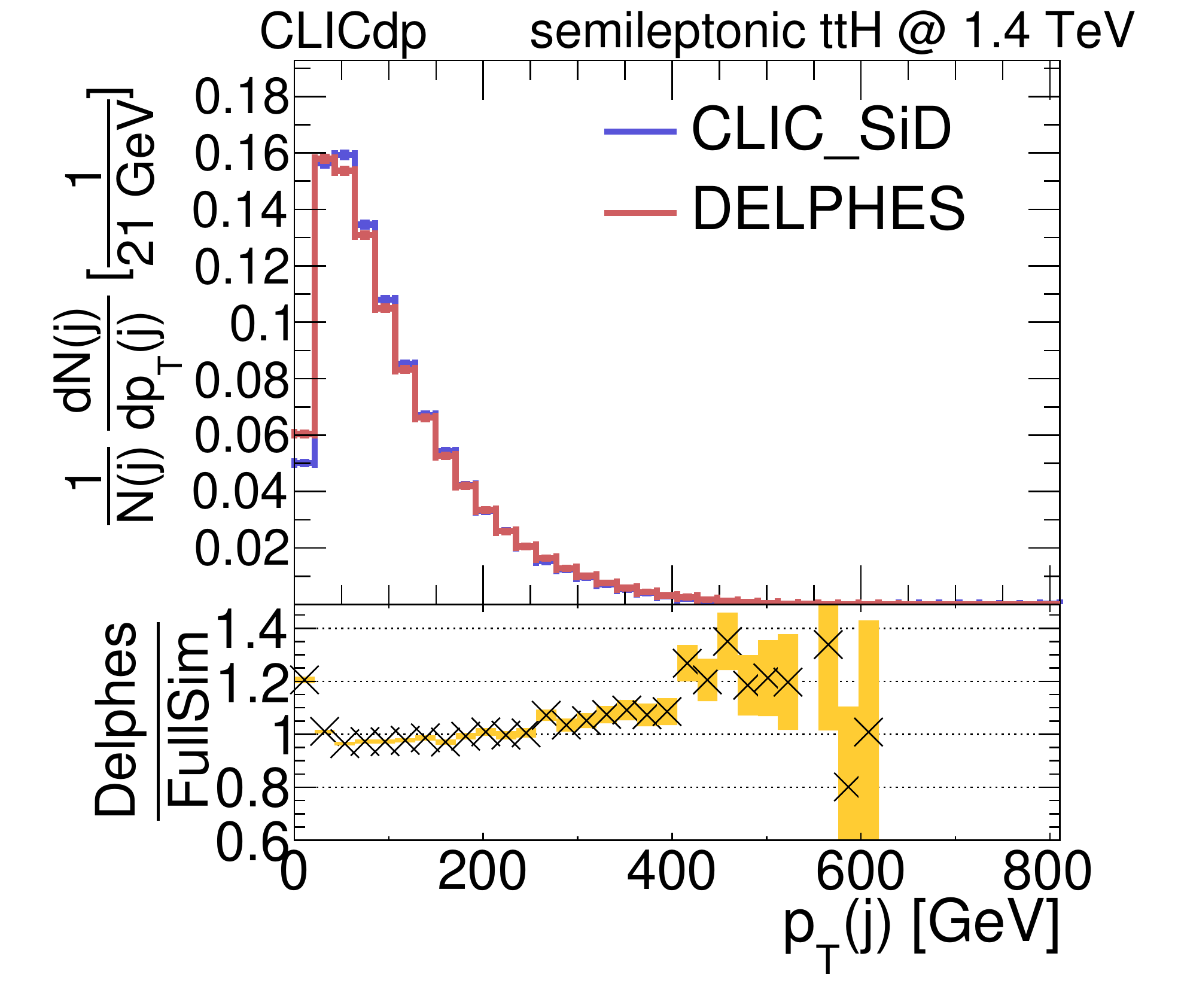}
      \includegraphics[width=0.495\textwidth]{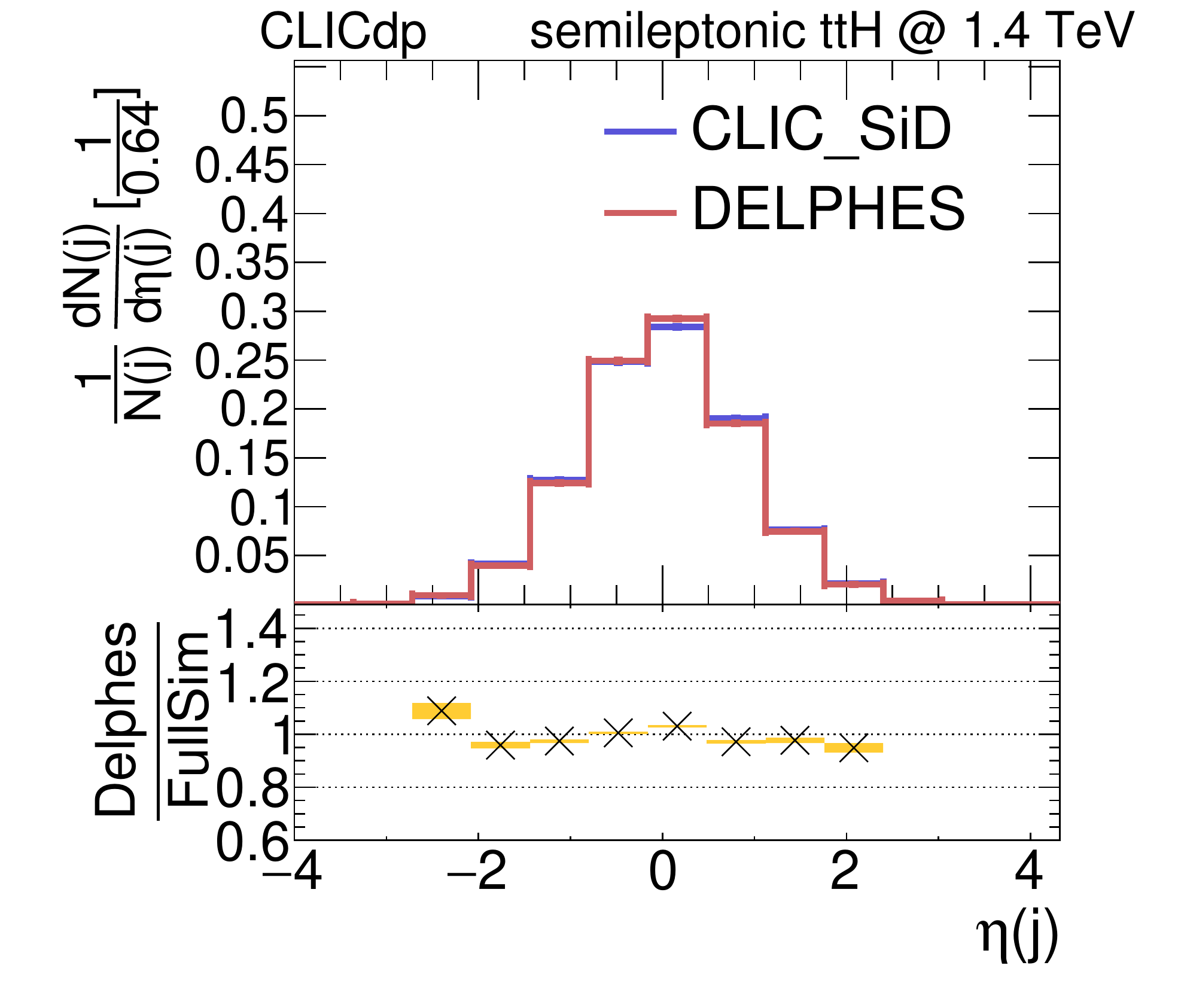}
  \caption[]{Comparison of
    jet transverse momentum (left) and pseudorapidity (right) in semileptonic $\ttbar \PH$ events for full
    simulation of CLIC\_SiD (blue) and \Delphes (red).}
      \label{fig:tthJets_app}
    \end{figure}

     The jet performance  is demonstrated in Fig.~\ref{fig:tthJets_app}. It
    includes the jet energy smearing mimicking the effects of \gghadrons background. The pseudorapidity distribution is
    well-modelled, while the \Delphes spectrum for transverse
    momentum shows an increase of events with low as well as higher
    transverse momentum.

\subsection{Di-boson production WW at 3\,TeV}\label{app:extraplots_ww}
\begin{figure}[ht]\centering
  \includegraphics[width=0.49\textwidth]{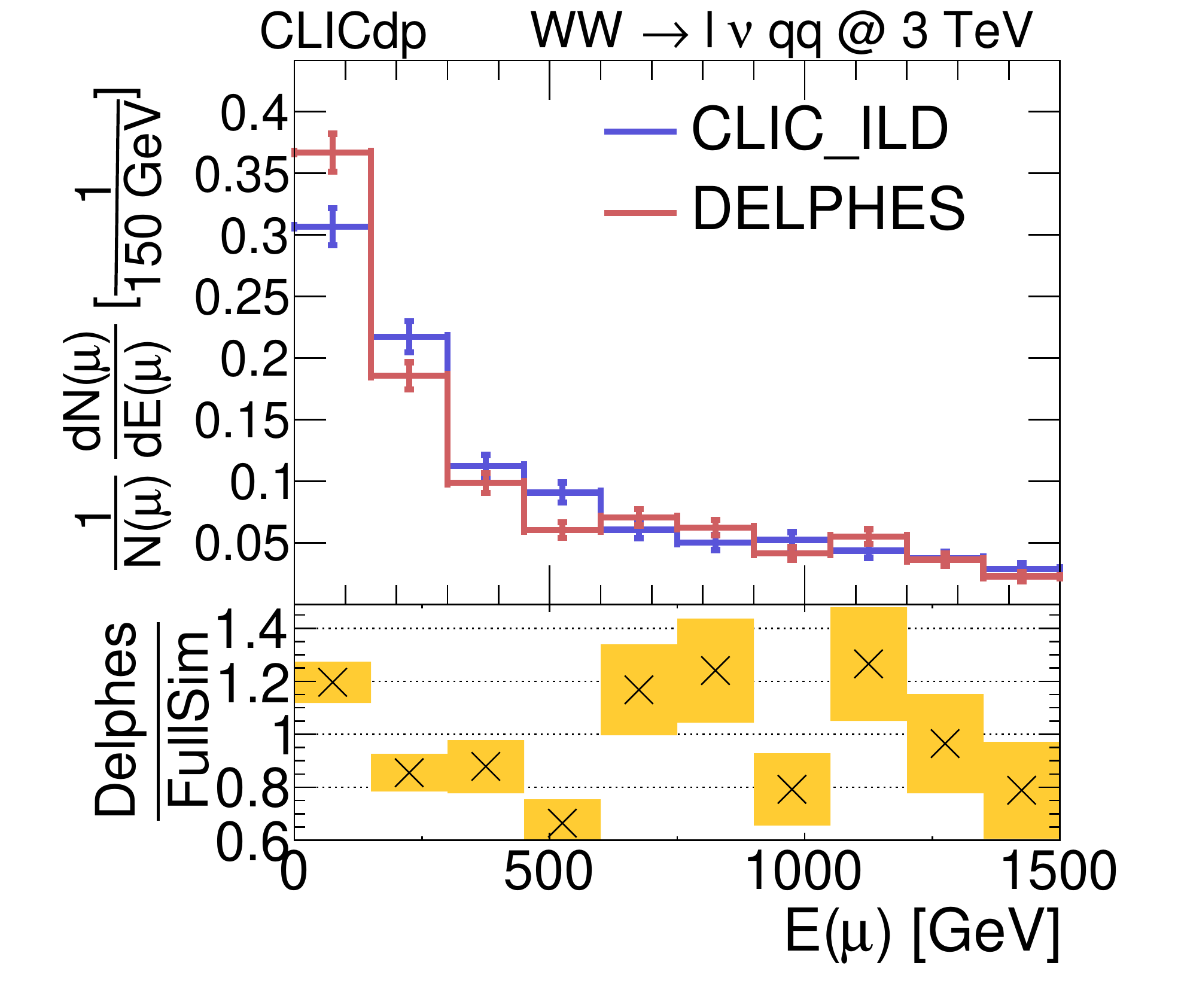}
  \caption[]{Validation of fast simulation performance for the muon
     energy for events with the final state $\ell \nu \qqbar$ at 3\,TeV
    with full simulation of CLIC\_ILD (blue) and fast simulation
    with \Delphes (red).}
  \label{fig:WWmuons_app}
\end{figure}

The muon energy distribution is validated comparing  \Delphes to full detector simulation with the CLIC\_ILD model as
shown in Fig.~\ref{fig:WWmuons_app}.

\begin{figure}[ht]\centering
 \includegraphics[width=0.49\textwidth]{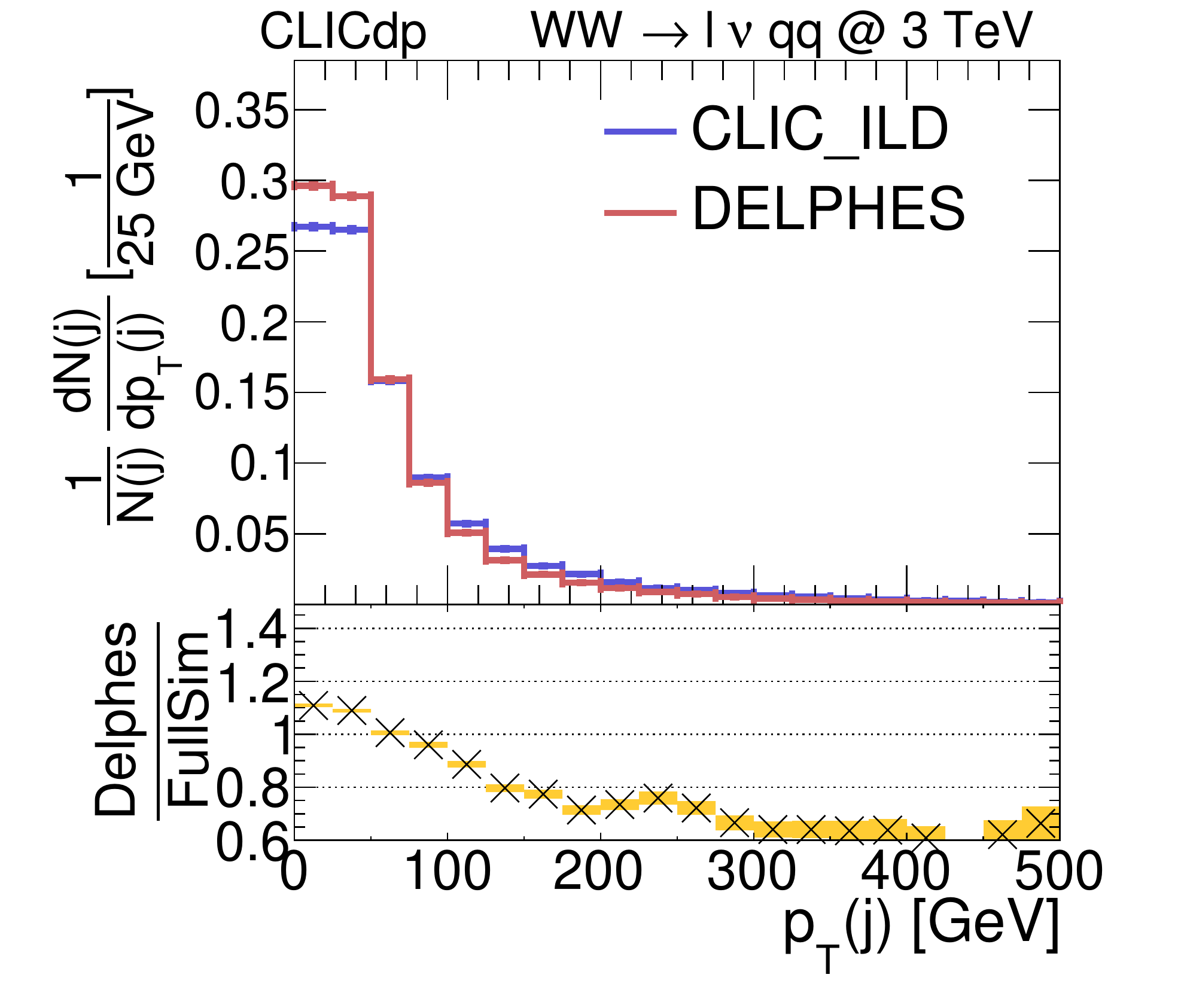}
    \includegraphics[width=0.49\textwidth]{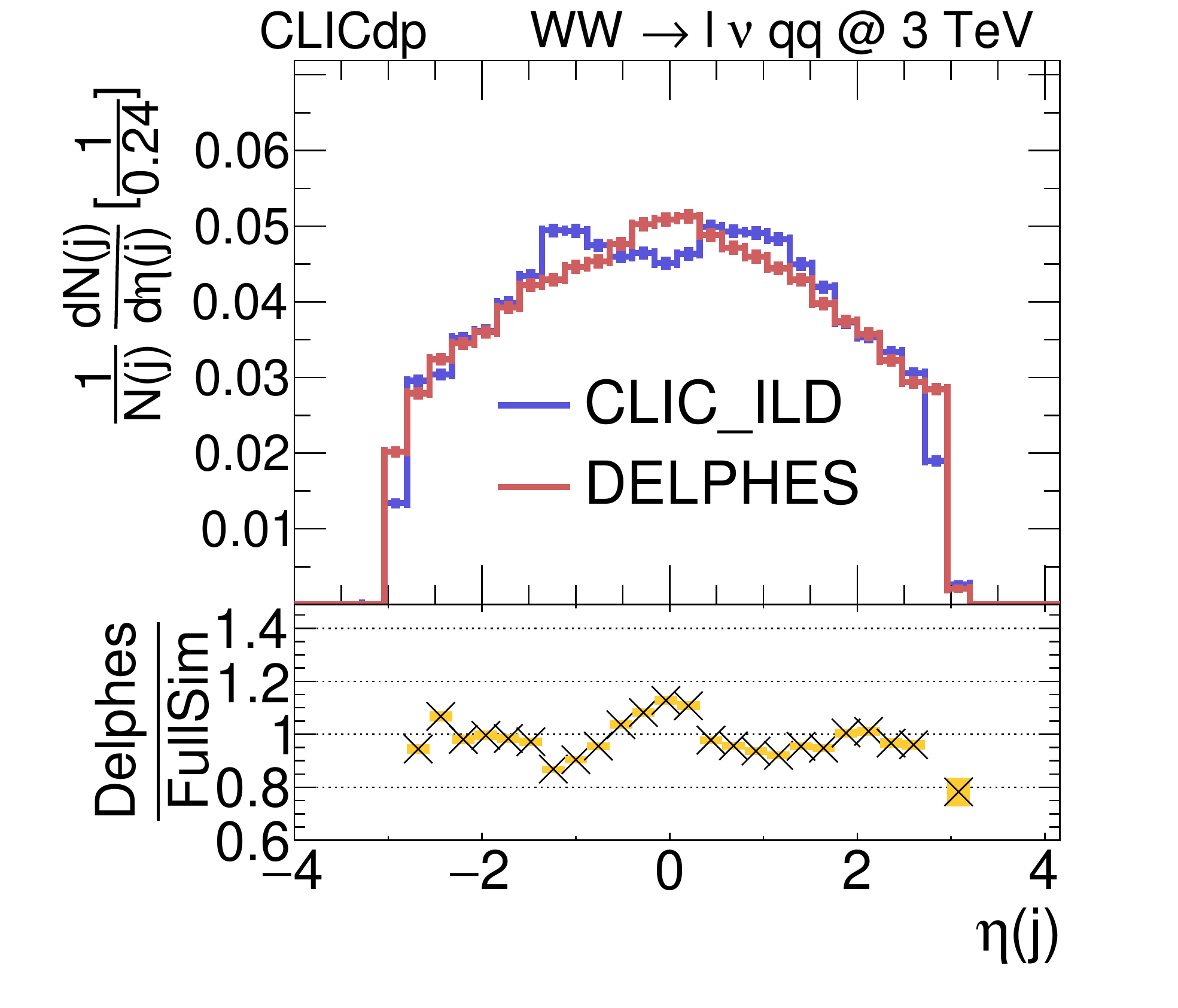}
    \caption[]{Validation of fast simulation performance for the
      transverse momentum of jets (left) and the pseudorapidity of jets
      (right)  using  events with the final state $\ell \nu \qqbar$ at 3\,TeV
      with full simulation of CLIC\_ILD (blue) and fast simulation
      with \Delphes (red).}
    \label{fig:WWjets_app}
  \end{figure}

 The  transverse momentum of the jets is 
  underestimated by the \Delphes simulation, and it does not show all
  features of the CLIC\_ILD detector geometry in the pseudorapidity,
  as shown in Fig.~\ref{fig:WWjets_app}.

\end{document}